\documentclass[preprint,11pt]{elsarticle}
\usepackage{lmodern} 
\usepackage[T1]{fontenc}
\usepackage[utf8]{inputenc}
\usepackage{amsmath}
\usepackage{amssymb}
\usepackage{tensor}
\usepackage{lineno}
\usepackage{color}
\usepackage{graphicx}
\usepackage{hyperref}

\DeclareMathOperator{\sign}{sign}

\journal{Computer Physics Communications}

\begin{document}

\begin{frontmatter}



\title{Modeling Cosmic-Ray Transport: A CRPropa based stochastic differential equation solver}


\author[a,b]{Lukas Merten\corref{author}}
\author[a,b]{Sophie Aerdker}

\cortext[author] {Corresponding author.\\\textit{E-mail address:} lukas.merten@rub.de}
\address[a]{Ruhr-Universität Bochum, Universitätsstraße 150, 44801 Bochum, Germany}
\address[b]{Ruhr Astroparticle and Plasma Physics Center (RAPP Center), Germany}

\begin{abstract}
We present a new code that significantly extends CRPropa's capabilities to model the ensemble averaged transport of charged cosmic rays in arbitrary turbulent magnetic fields. The software is based on solving a set of stochastic differential equations (SDEs).

In this work we give detailed instructions to transform a transport equation, usually given as a partial differential equation, into a Fokker-Planck equation and further into the corresponding set of SDEs. Furthermore, detailed tests of the algorithms are done and different sources of uncertainties are compared to each other. So to some extend, this work serves as a technical reference for existing and upcoming work using the new generalized SDE solver based on the CRPropa framework.

On the other hand, the new flexibility allowed us to implement first test cases on continuous particle injection and focused pitch angle diffusion. For the latter one we show that focused pitch angle diffusion leads to a drift velocity along the field lines that is defined by the fixed points of the pitch angle diffusion equation.
\end{abstract}


\end{frontmatter}

\section{Introduction}
\label{sec:intro}

The modeling of particle transport plays an important role in many different scientific fields. In this work, we focus on the description of the propagation of high energetic cosmic rays in the presence of turbulent background magnetic fields. However, the techniques described throughout this work could also be applied to different fields, from biology to neural computation, where diffusive transport or stochastic differential equations (SDEs) are of interest.

Whenever it is either computationally not possible or simply not necessary to model individual trajectories of charged particles, as it is done, e.g.\ in fully kinetic particle in cell (PIC) simulations, an ensemble averaged description of the collective motion is desired. Depending on the physical details that should be captured, different levels of averaging can be applied to the basic equation. 

In astroparticle physics one usually starts with the so called pitch angle transport equation, which is averaged over the gyro-phase as it can be neglected in most astrophysical scenarios. When transport perpendicular to the magnetic field line is ignored the focused pitch angle transport equation can be written as
\begin{align}
    \frac{\partial f}{\partial t} = \frac{\partial}{\partial \mu} \left(D_{\mu\mu} \frac{\partial f}{\partial \mu} \right)  - \frac{\partial}{\partial \mu} \left(\frac{(1-\mu^2)v}{2L(s)} f \right) - \frac{\partial}{\partial s}(\mu v f) + S \label{eq:focused_transport} \quad .
\end{align}
Here, $f$ is the distribution function, $\mu$ is the cosine of the pitch angle\footnote{The angle between the magnetic field line and the particle's momentum $\mu=\mathbf{B}\cdot\mathbf{p} / (Bp)$.}, $v$ is the particle speed, $D_{\mu\mu}$ is the pitch angle diffusion coefficient, $L$ is the focusing length, $S$ is a source/sink term and $s$ is the parameterization parameter of the field line. The focusing term accounts for phenomena like magnetic mirroring and depends on the gradient of the magnetic field strength $L(s) = - B(s) / \left( \partial B / \partial s \right)$.

Often the particle distribution can be assumed to be isotropic; meaning it follows a uniform distribution of $\mu$. In that case, one can average the transport equation over the pitch angle and gets the following transport equation
\begin{align}
    \frac{\partial f}{\partial t} = \nabla \cdot \left(\hat{\kappa} \nabla f - \mathbf{v}f \right) - \mathbf{w} \cdot \nabla f + \frac{p}{3} \nabla \cdot \mathbf{w} \frac{\partial f}{\partial p} + \frac{1}{p^2} \left( \frac{\partial}{\partial p} p^2 D_{pp} \frac{\partial f}{\partial p} \right) - Lf + S \label{eq:transport} \quad, 
\end{align}
including spatial and momentum diffusion and advection. Here, $f$ is now the pitch angle averaged distribution function, $\hat{\kappa}$ is the spatial diffusion tensor, $\mathbf{v}$ and $\mathbf{w}$ are two components of the advective flow (see below for the reason of this separation), $D_{pp}$ is the momentum diffusion coefficient, $L$ encodes losses or gains and $S$ describes again sources and sinks. Note, that in contrast to equ.~\ref{eq:focused_transport} the transport equ.~\ref{eq:transport} describes three dimensional spatial transport but is limited to one momentum or energy dimension. 

When the particle transport is non-Gaussian, usually classified by the nonlinear behavior of the mean squared displacement of particles $\langle \Delta x^2\rangle \propto t^{\zeta\neq 1}$, the transport equation becomes fractional. Here, we focus on superdiffusion with L\'evy flights, corresponding to the space-fractional transport equation
\begin{align}
\label{eq:space-fractional}
    \frac{\partial f}{\partial t} &= \kappa_{\alpha} \nabla^{\alpha} f  - u(x) \cdot \frac{\partial  f}{\partial x} - \frac{p}{3} \frac{\partial u}{\partial x} \frac{\partial  f}{\partial p} +  S(x, p, t) \quad , 
\end{align}
with the fractional dimension $\alpha = 2/\zeta$ and fractional diffusion scalar $\kappa_{\alpha}$ which here is assumed to be spatially constant. The fractional diffusion scalar has units depending on $\alpha$. The Riesz-derivative \citep{Gorenflo-etal-1999} is given by
\begin{align}
  \label{eq:riesz-riemann}
    \nabla^{\alpha} f(x) = -\frac{1}{2\cos(\alpha\pi /2)} (\tensor*[_{-\infty}]{D}{_x^{\alpha}} + \tensor*[_x]{D}{_{+\infty}^{\alpha}}) f(x) \quad ,
\end{align} 
with the Riemann-Liouville fractional derivative being defined as
\begin{align}
\label{eq:riemann-liouville}
   \tensor*[_0]{D}{_t^{1-\beta}} f(t) = \frac{1}{\Gamma(\beta)}\frac{d}{dt}\int_0^t(t-s)^{\beta-1}f(s)ds \quad.
\end{align}
Note, that we used the one dimensional fractional transport equation here as an example --- like most of the literature currently does (see e.g.~\cite{Metzler-Klafter-2004}) --- which can be generalized to more dimensions. Also momentum diffusion and additional loss processes have been neglected. 

We do not repeat the derivations of the essential equations introduced above\footnote{The interested reader is referred to e.g.~\cite{Schlickeiser-2002, Shalchi-2009} for the spatial diffusion, to e.g.~\cite{vandenBerg-etal-2020} for the focused pitch angle equation, and \cite{Metzler-Klafter-2000, Metzler-Klafter-2004} for the fractional transport equation.}, but rather want to rigorously derive the corresponding stochastic differential equations (SDEs) (see below) and explain the algorithms used to solve those SDEs (see sec.~\ref{sec:code}). This includes a discussion of different relevant sources of uncertainties and the validation of the most important parts of the software. Furthermore, we show detailed examples of some of the newly implemented features of the software such as momentum diffusion (see sec.~\ref{ssec:MomentumDiffusion}) and pitch angle diffusion (see sec.~\ref{ssec:pitch_angle}) and remind the reader of all other opportunities (diffusive shock acceleration (DSA), superdiffusion, parallel transport along arbitrary field lines, etc.) of the code that have been released earlier for completeness. 

\subsection{From Transport to Stochastic Differential Equations}
\label{ssec:FP2sde}
Every Fokker-Planck Equation (FPE)
\begin{align}
    \frac{\partial g(q_1, \dots, q_n, t)}{\partial t} = - \sum_{i=1}^n \frac{\partial}{\partial x_i} \left(A_i g \right) + \frac{1}{2} \sum_{i, j} \frac{\partial^2}{\partial x_i \partial x_j} \left(B_{ij} g \right) \quad, \label{eq:FP-forward}
\end{align}
where in our case the $q_i$ are usually the spatial and momentum coordinates, can be transformed into an equivalent set of stochastic differential equations (SDEs); It\^o's Lemma \cite{Ito-1951}.\footnote{We refer to the appendix \ref{apdx:distribution-function} for the derivation of the time backward case.} The corresponding SDE then reads:
\begin{align}
    \mathrm{d}\mathbf{x} = \Tilde{\mathbf{A}}\,\mathrm{d}t + \Tilde{\hat{B}}\,\mathrm{d}\mathbf{W}_t \label{eq:SDE} \quad ,
\end{align}
where $\mathrm{d}\mathbf{W}_t$ is a Wiener process and $\Tilde{\mathbf{A}}=\mathbf{A}$ is the deterministic drift and $\Tilde{\hat{B}}\Tilde{\hat{B}}^\dagger = 1/2 (\hat{B} + \hat{B}^\mathrm{t})$ describes the stochastic part of the motion. For the one dimensional case, or when the diffusion tensor is diagonal (i.e.\ parallel and perpendicular transport in local magnetic field coordinates), the stochastic term simplifies to $\Tilde{B_i} = \sqrt{B_i}$, see, e.g., \cite{Kopp-etal-2012} for descriptions how to derive the square root of the diffusion tensor if it is non-diagonal. The SDE (\ref{eq:SDE}) can be interpreted as the equation of motion for a phase-space element. Therefore, we call each realization of the solution of the SDE a trajectory of a \emph{pseudo-particle}.

As an example, we take a look at the transport equ.~\ref{eq:transport} and bring it into the form of a forward FP equation. Note, that the Helmholtz decomposition of the velocity field $\mathbf{u}=\mathbf{v} + \mathbf{w}$, where $\mathbf{w} = -\nabla \phi$ and $\mathbf{v} = \nabla \times \mathbf{A}$ was used. Only the wind component $\mathbf{w}$ has an influence on the adiabatic energy changes as $\nabla\cdot\mathbf{v}=\nabla\cdot(\nabla\times\mathbf{A})=0$. We start by transforming each summand into a form that is compatible with a FP equation.
\begin{align}
    \nabla \cdot (\hat{\kappa} \nabla f -\mathbf{v} f) &= \nabla^2(\hat{\kappa} f) - \nabla \cdot \left[(\nabla \hat{\kappa})f + \mathbf{v}f \right] \quad , \label{eq:trafo_kappa-f} \\
    -\mathbf{w} \nabla f &= -\nabla \cdot (\mathbf{w} f) + (\nabla \cdot \mathbf{w}) f \quad , \label{eq:trafo_w-f}\\
    \frac{p}{3} (\nabla \cdot \mathbf{w}) \frac{\partial f}{\partial p} &= \frac{\partial}{\partial p} \left(\frac{p}{3} (\nabla \cdot \mathbf{w}) f\right) - \frac{1}{3} (\nabla \cdot \mathbf{w})f \quad , \label{eq:trafo_adiabatic-f} \\
    \frac{1}{p^2} \left[\frac{\partial}{\partial p} \left(p^2 D\frac{\partial f}{\partial p} \right) \right] &= \frac{\partial^2}{\partial p^2} (Df) + \frac{\partial}{\partial p} \left[\left(\frac{2D}{p}- \frac{\partial D}{\partial p}\right) f\right] - \frac{\partial}{\partial p}\left(\frac{2D}{f}\right) f \quad . \label{eq:trafo_D-f}
\end{align}
Now inserting equs.~\ref{eq:trafo_kappa-f} - \ref{eq:trafo_D-f} into equ.~\ref{eq:transport} leads to
\begin{align}
    \frac{\partial f}{\partial t} &= \frac{1}{2} \nabla^2 (2\hat{\kappa}f) - \nabla \cdot\left[\left(\nabla\hat{\kappa} + \mathbf{v} + \mathbf{w} \right) f \right]  \notag \\ 
    &+ \frac{1}{2}\frac{\partial^2}{\partial p^2}(2Df) - \frac{\partial}{\partial p} \left[ \left( \frac{\partial D}{\partial p} - \frac{2D}{p} - \frac{p}{3}\nabla\cdot\mathbf{w} \right) f \right] \notag \\
    &-\left[-\frac{2}{3} \nabla\cdot\mathbf{w} + \frac{\partial}{\partial p} \frac{2D}{p}\right]f \quad , 
\end{align}
which leads to the following components of the SDEs
\begin{align}
    A_x = \left(\nabla\hat{\kappa} + \mathbf{v} + \mathbf{w} \right) \quad &, \quad A_p = \left(\frac{\partial D}{\partial p} - \frac{2D}{p} - \frac{p}{3}\nabla \cdot \mathbf{w} \right) \quad ,\\ \label{eq:SDE_coeff_f-forward}
    B^2_x = 2\hat{\kappa} \quad &, \quad B_p^2 = 2D \quad ,
\end{align}
where variables with subscribed $x$ refer to the three dimensional spatial and with subscribed $p$ to the one dimensional momentum SDE. Note that the last summand, proportional to $f$, is not yet taken into account. Generally, source and loss terms as in the following equation:
\begin{align}
    \frac{\partial f}{\partial t} = - Lf + S \quad , \label{eq:source-sinks}
\end{align}
cannot directly be included into the SDE but have to be treated by weighting the phase space elements or pseudo-particles in post-processing. Following the nomenclature of Kopp et al.~\cite{Kopp-etal-2012} we call the factor introduced by the loss terms ($-Ln$) \emph{path weight} $\omega$ and the one coming from the sources or sinks ($S$) is called \emph{path amplitude} $w$. 
The path weight for an individual time step is given by $\omega_i = \exp(-L(\mathbf{r}_i, p_i, t_i)\Delta t_i)$. Since these weights are multiplicative this leads to:
\begin{align}
    \omega_j =  \exp \left(-\sum_{i=0}^j L(\mathbf{r}_i, p_i, t_i)\Delta t_i\right) \quad.
\end{align}
Analogously the path amplitude $w$ is increased (decreased) if a particle encounters a source (sink) region $w_i = w_{i-1} + S_i \omega_i \Delta t_i$, which can be written as:
\begin{align}
    w_i = \sum_{j=0}^{i} S_j \Delta t_j  \exp \left(-\sum_{k=0}^j L(\mathbf{r}_k, p_k, t_k)\Delta t_k\right) \quad .
\end{align}

From that we can derive the additional path weight which is an essential part to derive the correct distribution function
\begin{align}
    \omega_\mathrm{transp., i} &= \exp\left[ \left(\frac{2}{3} \nabla\cdot\mathbf{w} - \frac{\partial}{\partial p} \frac{2D}{p}\right) \Delta t_i \right] \notag \\
    &= \exp\left[ \left(\frac{2}{3} \nabla\cdot\mathbf{w} - \frac{2}{p} \left(\frac{\partial D}{\partial p} - \frac{1}{p} \right) \right) \Delta t_i \right] \quad . \label{eq:transport_weight-f}
\end{align}

The derivation of the time backward case (see section \ref{apdx:distribution-function}) and for the particle number density $n=fp^2$ (see section \ref{apdx:number-density}) can be found in the appendix. It has to be noted, that the parameters of the SDEs and the weighting terms differ for all four equations. 

Similarly, the equations for the focused pitch angle transport can be derived (see section \ref{apdx:pitchangle}) and read:
\begin{align}
    \mathrm{d}s &= v\mu\,\mathrm{d}t \notag \\
    \mathrm{d}\mu &= \left(\frac{v}{2L}(1-\mu^2) + \frac{\partial D_{\mu\mu}}{\partial \mu}\right)\,\mathrm{d}t + \sqrt{2D_{\mu\mu}}\,\mathrm{d}W_t \label{eq:sde_pitchangle} \quad .
\end{align}
Note, that the transport along the field line is modeled with an ordinary differential equation, where the stochastic behavior is only indirectly included through coupling with the diffusive pitch angle. This leads to different pseudo-particle trajectories when comparing spatial diffusion with pitch angle diffusion models to each other (see section \ref{ssec:pitch_angle}.

\subsection{Derivation of physical quantities}
\label{ssec:sde2physics}

Each solution of the SDE can be interpreted as one possible realisation of the time evolution of the distribution function $f$ or the number density $n$, respectively. This means that the simulation output is usually a table of observations of pseudo-particle properties, such as position, energy, particle type, etc. To derive a result that is compatible with physical observations these individual solutions have to be averaged over. Usually this is done by creating histograms of the relevant quantities, e.g., the modeled pseudo-particle energy to derive the energy spectrum. The energy spectrum can be approximated by
\begin{align}
    \frac{\mathrm{d}n}{\mathrm{d}E} (E) = \sum_i \frac{W_i}{\Delta E(E)} \quad , 
\end{align}
where all particle in the energy range $E_i\in(E\pm\Delta E/2)$ are counted. The weight $W$ includes the path weight $w_i$, the path amplitude $\omega_i$, and any additional weighting term of each pseudo-particle. These additional weights can, e.g., normalize the spectrum to physical units (see, e.g., \cite{Merten-etal-2018}) or introduce a re-weighting of the simulation result. 

Re-weighting is a technique to change some\footnote{Re-weighting is only possible for those parts of the total parameter space that have been probed by the ensemble of simulated pseudo-particles. This means, e.g.~ that the ratio of the strength of two sources can be changed after the simulation. Of course a third, not simulated source, cannot be included through re-weighting.} parameters of the injected simulation values. The most common example is the re-weighting of the simulated particle energy spectrum. To achieve equal statistics in each energy bin, simulations are usually run with a simulated spectral index $s_\mathrm{inj}=-1$, where $\mathrm{d}n/\mathrm{d}E\propto E^s$. The weights to mimic a modeled spectral index of $s_\mathrm{mod}$ is given by $w_\mathrm{mod}=E_0^{s_\mathrm{mod} - s_\mathrm{inj}}$, here $E_0$ is the energy of the pseudo-particle at the source. Note, that this weight does not conserve the number or energy density which can, however, be achieved with more complex weighting schemes.

The minimal uncertainty of such approximations of physical quantities are then given by the Monte Carlo error. This uncertainty for a property $X$ can be defined as:
\begin{align}
    \Delta = X / \sqrt{N} \quad ,
\end{align}
where $N$ is the number of independent observations that have been used in the averaging process, e.g., the number of pseudo-particles per energy bin.

Section \ref{ssec:splitting} and section \ref{ssec:cont_injection} give more information on how to evaluate the independence of observations and how to derive stationary approximations from time evolving simulations.

\section{Implementation}
\label{sec:code}

With CRPropa~3.1 \cite{Merten-etal-2017} the open source propagation framework was extended by a simple solver for the ensemble averaged description of the cosmic-ray number density. Instead of discretizing the transport equ.~\ref{eq:transport} on a grid it is transformed into a set of SDEs. This approach is complementary to other software tools like Galprop \cite{Strong-Moskalenko-1998}, DRAGON \cite{Evoli-etal-2008, Gaggero-etal-2013}, or PICARD \cite{Kissmann-2014, Kissmann-etal-2015}. The previously implemented solver \texttt{DiffusionSDE} was optimized for Galactic transport \cite{Merten-etal-2017}. The adaptive fieldline integrator allows to solve the SDEs in arbitrary coherent magnetic background fields, e.g.\ in large scale structure Galactic field models like \cite{Jansson-Farrar-2012, Ferriere-Terral-2014, Kleimann-etal-2019, Unger-Farrar-2024}. The main limitation of the previous implementation is the assumption of spatially constant eigenvalues of the diffusion tensor. Also only the new version allows to solve for other transport equations than the number density one, such as the pitch angle diffusion equation or solving for the distribution function $f$ by calculating weights.

The new implementation of a generalized solver for SDEs still focuses on high energy particle transport, assuming the ultra-relativistic limit ($v=c$). The spatial transport equation~(\ref{eq:transport}) also assumes an isotropic momentum component of the distribution function $f(\mathbf{r}, \mathbf{p}, t)=f(\mathbf{r}, p, t)$. However, it allows for spatially varying diffusion coefficients making it necessary to implement drift terms induced by derivatives of the diffusion tensor (see $\nabla\hat{\kappa}$ in equ.~\ref{eq:SDE_coeff_f-forward}). The pitch angle diffusion equation can be solved for anisotropic distributions $f(s, \mu, t)$, too. 

The current implementation separates the setup into different modules for physics definitions (\texttt{DiffusionTensor}), the transformation into SDEs (\texttt{SDEParameter}), and the solver \texttt{SDESolver}. 

Furthermore, the \texttt{SDEsolver} is extended to take anomalous diffusion into account by changing the driver of the stochastic process. This way, the space-fractional FPE is solved by integrating the corresponding SDEs. The Wiener process for Gaussian diffusion is changed to a L\'evy process for superdiffusion considering L\'evy flights. With the modular structure the implementation of subdiffusive processes including \emph{waiting times} is straightforward but remains subject for a future version. 

The new structure also allows for time-dependent background fields, diffusion coefficients and injection of particles. With the new \texttt{SDESolver}, position and energy/momentum are integrated in the same step and not after each other. This is more accurate especially when time-dependent background fields are taken into account.

\subsection{DiffusionTensor}
\label{ssec:diffusiontensor}

The \texttt{DiffusionTensor} module is the base class for the physical model of the diffusion tensor. It provides functions to calculate the spatial and momentum diffusion coefficients and their derivatives. The diffusion tensor can be defined either in the laboratory frame or in the local frame of the coherent magnetic field line. In the latter case the $x$-, $y$-, and $z$-components correspond to the tangential, normal, and binormal direction of the local trihedron of the magnetic field line (see sec.~\ref{ssec:fieldline}). 

\subsection{SDEParameter}
\label{ssec:sdeparameter}
The \texttt{SDEParameter} module calculates the deterministic $A$ and stochastic coefficients $B$ and weights ($w$ and $\omega$) that are passed to the solver module. As explained in section \ref{sec:intro} the transport equation, e.g.\ for number density $n$ or distribution function $f$, makes a difference in the exact transformation from transport equation, e.g., equ.~\ref{eq:focused_transport}, 
to SDE, e.g.~\ref{eq:sde_pitchangle}. Therefore, one class for each case is implemented. 

If the only input to the \texttt{SDEParameter} is a \texttt{DiffusionTensor} class the SDE parameters will be calculated in the lab frame. Providing additionally a \texttt{MagneticField} class will subsequently solve the diffusion part in the frame of the magnetic field line (see below for details), where the $x$-axis is associated with the parallel direction and the $y$-$z$ plane is associated with the two perpendicular directions.
 
\subsection{SDESolver}
\label{ssec:sdesolver}
The \texttt{SDESolver} module solves the SDE and updates the pseudo-particles' positions and energies. The default case solves the SDE with an Euler-Mayurama scheme 
\begin{align}
    q_{n+1} = q_n + A_qh + B_q \eta_{n, q} \sqrt{h} \quad , \label{eq:SDE_Lab}
\end{align}
where $q$ is either a component of the spatial position $x_i$, the absolute momentum $p$, position along the field line $s$ or the pitch angle $\mu$ of a phase-space element, $h$ is the time step, and $\eta_{n, q}$ is a normally distributed random number with unit variance and vanishing mean. For a discussion on the stability of the algorithm the reader is referred to, e.g.~\cite{Gardiner-2009, Merten-etal-2017, Merten-etal-2018, Achterberg-Schure-2010}.
When the diffusion process is defined in the frame of the local coherent background field the integration of the spatial part of the SDE can be written as
\begin{align}
    \mathbf{x}_{n+1} = \mathbf{x}_{n} + \sum_i A_i \mathbf{e}_i h + \left( B_\parallel \eta_\parallel \mathbf{e}_\parallel + B_{\perp, 1} \eta_{\perp, 1} \mathbf{e}_{\perp, 1} +B_{\perp, 2} \eta_{\perp, 2} \mathbf{e}_{\perp, 2} \right) \sqrt{h} \quad , \label{eq:SDE_fl}
\end{align}
where the deterministic step is in the lab frame with the unit vectors $\mathbf{e}_i$ in $x, y, z$-direction. The stochastic step is calculated in the frame of the local field line, with $\{\mathbf{e}_\parallel, \mathbf{e}_{\perp, 1}, \mathbf{e}_{\perp, 2}\}$ being the parallel and the two perpendicular directions of the field line. Technically, the frame of the local field line is approximated by the tangential, normal, and binormal vector of the magnetic field line at position $\mathbf{x}_n$ (see sec.~\ref{ssec:fieldline}). The solver can optimize the length of the next time integration step $h$ based on the accuracy of the current field line integration (see \ref{apdx:fli} for more information). 

\subsubsection{Pitch Angle Diffusion}
\label{sssec:pitchanglediffusion}

The solver for focused pitch angle transport looks very similar
\begin{align}
    \mu_{n+1} &= \mu_n + A_\mu h + B_\mu \eta_\mu \sqrt{h} \\
    s_{n+1} &= s_n + v\mu_n h \quad , 
\end{align}
where the coefficients $A_\mu$ and $B_\mu$ are derived in the appendix (see \ref{apdx:pitchangle}). Note, that in contrast to the spatial diffusion equation, the pitch angle $\mu$ is limited to a finite regime $|\mu|\leq 1$. To ensure this, reflective boundaries are applied during the propagation step (see \ref{sapdx:pitch_angle_boundary} for the details).

When a non-uniform magnetic background field is used the new pseudo-particle positions are calculated by integrating the magnetic background field line for a distance $L=v\mu h$ (see equ.~\ref{eq:fieldline}). This allows to perform pitch angle diffusion studies in complex magnetic field structures, such as the Solar magnetic field model as, e.g.\ given by Parker \cite{Parker-1958}

\subsubsection{Superdiffusion}
\label{sssec:superdiffusion}
Superdiffusive transport following the space-fractional FPE (\ref{eq:space-fractional}) is implemented by applying the generalized It\^o's lemma \citep{Ito-1951, Magdziarz-Weron-2007}. In the SDE, the Wiener process is changed to a L\'evy process, which is proportional to $\mathrm{d}t^{1/\alpha}$. Furthermore, the stochastic component of the SDE is scaled differently: $B_q^2 = 2 \kappa_{\alpha}^{2/\alpha}$ \footnote{Note, that $\zeta$, the anomalous diffusion exponent (see sec.\ref{sec:intro}) is equal to $2/\alpha$.}. This leads to a small change of the algorithm given in equ.~\ref{eq:SDE_Lab}:
\begin{align}
    q_{n+1} = q_n + A_qh + B_q \eta_{n, q} h^{1/\alpha} \quad , \label{eq:SDE_Lab_Levy}
\end{align}
where $\eta$ is drawn from an alpha-stable L\'evy distribution (see sec.~\ref{ssec:Superdiffusion}). The \texttt{SDEParameter} module provides the scaled component $B_q$. Note, that the fractional diffusion tensor has different units, $[\kappa_\alpha] = \mathrm{m}^\alpha / \mathrm{s}$.

\subsubsection{Adaptive Timestep for DSA}
\label{sssec:adaptive_dsa}
From the transport equ.~\ref{eq:transport} follows, that the advection has to be differentiable. For modeling DSA this implies a finite shock width. To properly resolve it, so that modeling DSA leads to the analytically predicted spectra the deterministic and stochastic steps have to fulfil the following inequality (see, e.g., \citep{Aerdker-etal-2024, Achterberg-Schure-2010, Kruells-Achterberg-1994})
\begin{align}
    \left( \left( \nabla \cdot \hat{\kappa} + \mathbf{u} \right) h \right) \cdot \mathbf{e}_{sh} \leq l_\mathrm{sh} \lesssim \sqrt{2} \left( \hat{\kappa} \mathbf{\eta} \right)  \cdot \mathbf{e}_{sh} \label{eq:Achterberg}
\end{align}

An adaptive step for DSA can be added to the \texttt{SDEEMSolver} which scales the next step according to the current position and energy of the candidate. For that, the minimal and maximal time steps are calculated according to the inequality given by Kruells\&Achterberg, 1994 \cite{Kruells-Achterberg-1994}. The maximal time step limits the advective step --- determined by the speed of the background flow and drift due to a spatially dependent diffusion tensor --- to be smaller than the chosen shock width. The minimal time step ensures that the diffusive step --- a measure for the stochastic step that depends on the (energy-dependent) diffusion coefficient --- is larger than the shock width $l_{\mathrm{sh}}$. Only when the inequality is fulfilled, acceleration at the shock is efficient and leads to the expected results. 

The DSA adaptive step assumes a spherically symmetric shock and is limited to a region $r_{\mathrm{sh}} \pm \Delta r$ around the shock. Outside this region, the next step is limited to the radial distance to the shockradius. This prevents overshooting the shock region and with that, the adaptive range can be held small ($\Delta r$ still depends on the shock width, chosen wind speed and diffusion tensor). In sec.~\ref{sssec:redsa} the adaptive step is used to limit the step size close to the shock. All that reduces the computation time while keeping the accuracy high enough to resolve the shock properly.


\subsection{Candidate Splitting}
\label{ssec:splitting}
The acceleration time scale of DSA is usually very large compared to the integration time step, especially when equ.~ \ref{eq:Achterberg} needs to be fulfilled. Furthermore, DSA leads to a rather steep energy spectrum --- fewer particles the higher the energy is. This results in an intrinsic worsening of the statistical uncertainty with increasing energy. Simply simulating more pseudo-particles can generally deliver any required statistic at some point but wastes computing time on low energetic particles. 

One way around this technical problem is importance sampling by the splitting of pseudo-particles. Here, a \texttt{Candidate} is split into two new \texttt{Candidates} when a certain criterion is met. When modeling particle acceleration this criterion is usually defined as some sort of energy threshold, e.g., when the pseudo-particle's energy has doubled since the last split. To conserve the total energy and particle number the \texttt{Candidate's} weight has to be adjusted by $w=n_\mathrm{split}^{-1}$, where $n_\mathrm{split}$ is the number of injected \texttt{Candidates}. This weight has to be taken into account in any further analysis, e.g., when the number density or energy spectra are derived.

\begin{figure}[htbp] 
    \centering 
    \includegraphics[width=.49\textwidth]{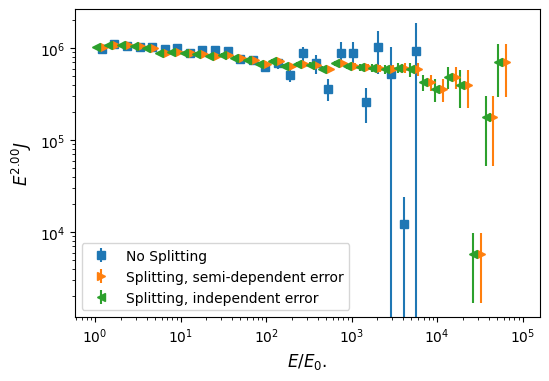}
    \hfill
    \includegraphics[width=.49\textwidth]{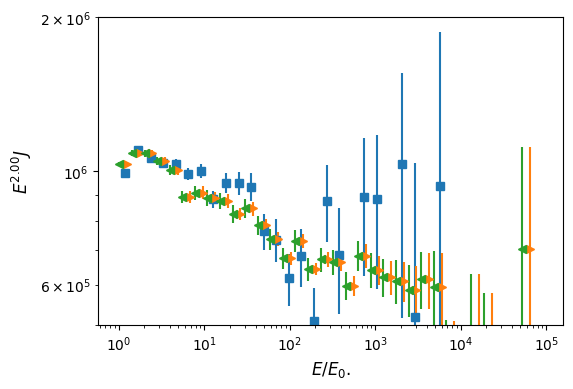}
    \caption{Left: Spectra obtained with and without using the \texttt{CandidateSplitting} module at simulation time $t = 100\;t_0$ evaluated at the shock. The shock compression ratio is $q=4$ and both simulations reproduce the expected spectral slope $s=-2$. The simulation with candidate splitting reaches higher energies up to $10^5\;E_0$ with smaller errors. Two different methods to evaluate the errors when CandidateSplitting is used are compared. Right: Close up to the stationary solution. }
    \label{fig:CandidateSplitting}
\end{figure}

Figure \ref{fig:CandidateSplitting} shows a comparison for modeled energy spectra from diffusive shock acceleration with and without applying the particle splitting. In both simulations $N = 10^4$ pseudo-particles are injected. With the \texttt{CandidateSplitting} the number grows during simulation time by a factor of $1.5$. Note, that the increase in runtime is less than that, since \texttt{Candidates} are added during simulation time. It is clearly visible that at the highest energies much better agreement with the analytical expectation is reached when particle splitting is included. 

Pseudo-particles created during the simulation depend to some extent on their \emph{parents}. This has to be taken into account when uncertainties are estimated. Figure \ref{fig:CandidateSplitting} compares errors obtained by treating all newly created pseudo-particles as instances of their parents or as individual pseudo-particles. The first \emph{independent} error is an overestimation of the uncertainty, since the new \texttt{Candidates} develop independently of their parents after creation. The second \emph{semi-dependent} error  is an underestimation of the uncertainty. Figure \ref{fig:CandidateSplitting} shows, that the difference in the uncertainty is negligible.
Both error estimations take the dependency of pseudo-particles due to the integration in time into account (see sec.~\ref{ssec:cont_injection} for a detailed discussion). 

Note, that particle splitting should not be applied for processes that result in very hard energy spectra, such as some cases of momentum diffusion. Where more particles are naturally expected at high energies the current implementation of particle splitting only leads to higher computing times without significantly impacting the statistics.
    
\subsection{Continuous Injection}
\label{ssec:cont_injection}

With time-dependent background fields, we also allow for continuous injection of pseudo-particles during simulation time. Up to now, all pseudo-particles were injected at $t = 0$ and propagated until the simulation stopped given by the individual boundary conditions. Since the SDE approach is independent on the source function and can be re-weighted in the later analysis, continuous injection of particles was realized by integrating \emph{snapshots} of the solution made during the simulation. This Green's function ansatz is discussed in detail in \cite{Merten-etal-2018} and, specifically for diffusive shock acceleration in \cite{Aerdker-etal-2024}. 

Here, we compare solutions obtained with continuous injection of $N$ \texttt{Candidates} with the time integration of $N_{\mathrm{t}}$ \texttt{Candidates} over $N_{\mathrm{obs}}$ observations. The use of time integration is faster since less pseudo-particles ($N_{\mathrm{t}} < N$) have to be injected to reach the same density of the phase space sampling. However, two uncertainties come with that advantage: Due to the integration in time, the same pseudo-particle can be in the same bin multiple times. And, the integration in time depends on the resolution $\Delta T$ of the snapshots taken during simulation, which is usually chosen larger than the simulation time step $h$. 

\paragraph{Number Density}
We evaluate these uncertainties by integrating the diffusion-advection equation
\begin{align}
    \mathrm{d}x = u \,\mathrm{d}t + \sqrt{2 \kappa} \,\mathrm{d}W_t,
\end{align}
with constant advection $u$. \texttt{Candidates} are injected at $x = 0$ and advected with positive $u$ towards $x \rightarrow \infty$. Few make it back against the flow. A number density profile similar to the one at a shock develops. The analytical steady state solution is given by
\begin{align}
    n_\mathrm{Gauss}(x, u, \kappa) = \frac{N_0}{u} \left(\exp^{\frac{|u|}{\kappa x}} \left(1 - \Theta(x)\right) + \Theta(x)  \right) \label{eq:stat_gauss} \quad ,
\end{align}
where $\Theta(x)$ is the Heaviside step function and $N_0=N N_\mathrm{obs} / T_\mathrm{max}$ is a normalization factor. For comparison also a linear ($n_\mathrm{linear}(x, a, b)=ax + b$) and a constant ($n_\mathrm{constant}(x, c)=c$) fit have been performed for the downstream region ($x>0$). 

The simulations were done with normalized units of $u=1\,\mathrm{ms}^{-1}$, $\kappa=1\,\mathrm{m}^2\mathrm{s}^{-1}$, and a simulation time step $h=10^{-3}\,\mathrm{s}$. The time evolution simulation model used $N_t=1000$ pseudo-particles and $N_\mathrm{obs}=250$ observations that are done at intervals of $\Delta T = 250 h$. In addition to the continuous simulation with $N=2.5\times10^5$ pseudo-particles, a model based on 250 individual simulations was computed. Here, the simulation time was increased by $\Delta T$ for each run and 1000 candidates were modeled per run. This model is very similar to the time evolution run, but the candidates for each observation time are truly independent of each other.

\begin{figure}[htbp]
    \centering
    \includegraphics[width=1\linewidth]{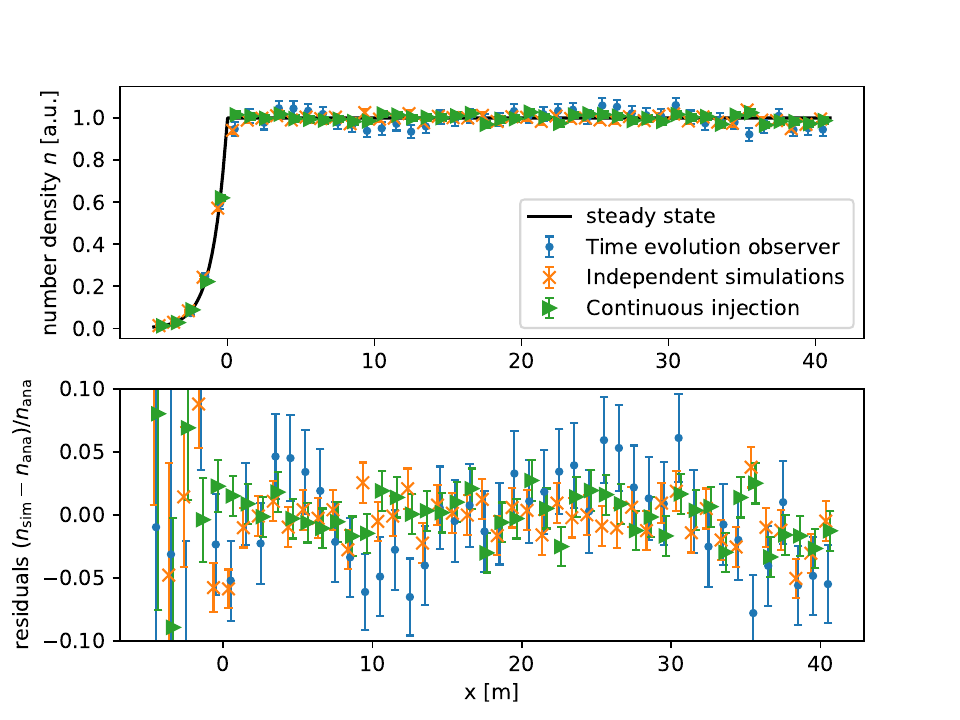}
    \caption{Number density in a diffusion advection model. Different ways to approximate the stationary solution (black solid line) are compared to each other --- time evolution observer (blue), independent simulation (green) and continuous source injection (green). Each simulation result contains the same number of candidates in the final data set.}
    \label{fig:error estimation}
\end{figure}

The number density $n$ as shown in fig.~\ref{fig:error estimation} and used for the fits is calculated as the number of candidates per bin divided by the bin width. To approximate the time stationary solution best, all candidates, including all observed trajectory length (in case of the continuous or time evolution models) or all individual runs (in case of the individual simulations) are used. The uncertainty is estimated as the Monte Carlo error so that the final approximation of the stationary solution is
\begin{align}
    n_\mathrm{sim}(x) = (N \pm \sqrt{\Tilde{N}}) / \Delta x \quad ,
\end{align}
where $N$ is the number of pseudo particles with $x_i\in(x-\Delta x/2, x+\Delta x/2)$, $\Delta x$ is the bin width, and $\Tilde{N}$ is the number of candidates used for the error estimation. In our new, more conservative evaluation of the uncertainties, this is given by the number of individual candidates --- differentiated by their unique serial number. In older publication we often used the total number of observations $N=\Tilde{N}$, where the same candidate could have been observed at different times in the same spatial bin, for the uncertainty estimation. Naturally, the new uncertainty estimates are always larger or equal to the old ones as $\Tilde{N}\leq N$. 

Figure \ref{fig:error estimation} shows the simulation results for the three different approaches to model the stationary solution (see equ.~\ref{eq:stat_gauss}) of the diffusion-advection scenario. It can be seen that all approaches lead to good estimates of the solution independent of the position. The lower panel shows the residuals of the simulations when compared to the stationary solution. The difference in the larger error bars for the time evolution model (blue) comes from the fact that it contains less independent observations than the other two models, also the fluctuation are larger. 

The qualitative observations from above are confirmed when looking at the fitting results (see tab.~\ref{tab:fit_parameters}). It can be seen that especially the constant region is well modeled in all three approaches but the continuous injection performs better for the modeling of the region $x<0$ (lower $\chi^2$ values for the Gauss fit). Note, that as expected the best fit results for the time evolution model do not deviate depending on the uncertainty estimation method. However, the goodness of fit increases for the new evaluation of the uncertainty.

\begin{table}[htbp]
    \centering
    \caption{Fitting results. Comparing the goodness of the fits of three different approaches to model the stationary solution of a diffusion advection scenario. Parameter 1 is either $u$, $c$ or $a$ according to equ.~\ref{eq:stat_gauss} and following lines and Parameter 2 is $\kappa$ or $b$.}
    \label{tab:fit_parameters}
    \begin{tabular}{l|l|rrr}
         & \textbf{Fit} & \textbf{Parameter 1} & \textbf{Parameter 2} &  $\mathbf{\chi^2/N_\mathrm{dof}}$ \\ \hline
         \textbf{Time Evolution} &&&& \\   
         --- new error &&&& \\  
         & Gauss & $1.005 \pm 0.028$ & $1.007 \pm 0.006$ & 1.29 \\
         & constant & $0.994 \pm 0.006$ & -- & 1.3  \\
         & linear & $(-5 \pm 5) \times 10^{-4}$ & $1.005 \pm 0.012$ & 1.30 \\
         \textbf{Independent} &&&& \\ 
         & Gauss & $1.017 \pm 0.016$ & $1.006 \pm 0.003$ & 1.58 \\
         & constant & $0.9964 \pm 0.0026$ & -- & 1.08  \\
         & linear & $(-3.5 \pm 2.2) \times 10^{-4}$ & $1.003 \pm 0.005$ & 1.04 \\
         \textbf{Continuous} &&&& \\ 
         & Gauss & $1.008 \pm 0.013$ & $1.0006 \pm 0.0025$ & 1.06 \\
         & constant & $0.9988 \pm 0.0026$ & -- & 1.08  \\
         & linear & $(-2.9 \pm 2.2) \times 10^{-4}$ & $1.005 \pm 0.005$ & 1.06 \\
         \textbf{Time Evolution}  &&&& \\ 
         --- old error &&&& \\
         & Gauss & $1.006 \pm 0.029$ & $1.006 \pm 0.006$ & 5.41 \\
         & constant & $0.995 \pm 0.006$ & -- & 5.48  \\
         & linear & $(-5 \pm 5) \times 10^{-4}$ & $1.006 \pm 0.012$ & 4.47 \\
    \end{tabular}
\end{table}

We conclude that although the uncertainties have probably been underestimated in some older publication using the \texttt{DiffusionSDE} module of CRPropa, the best fit values of any tested model are very likely correct, nevertheless. However, we recommend to use the new, based on truly independent observations, uncertainty estimates in the future. In some cases, namely when \texttt{CandidateSplitting} is used, the fully independent uncertainty estimation, using only the number of unique origin serial numbers per bin for $\Tilde{N}$, overestimates the true uncertainty, as splitted candidates follow independent trajectory after they have been split (see also sec.~\ref{ssec:splitting}).

\paragraph{Energy Distribution}
We further explore the uncertainties by looking at a second simulated quantity that depends on the position of the pseudo-particles, their energy $E$. With the advection $u$ modeling a shock profile, abruptly slowing down at $x = 0$, \texttt{Candidates} gain energy when they cross the shock. An example of DSA is given in sec.~\ref{sec:DSA}, for details we refer to \cite{Aerdker-etal-2024}. 

\begin{figure}[htbp]
    \centering
    \includegraphics[width=1\linewidth]{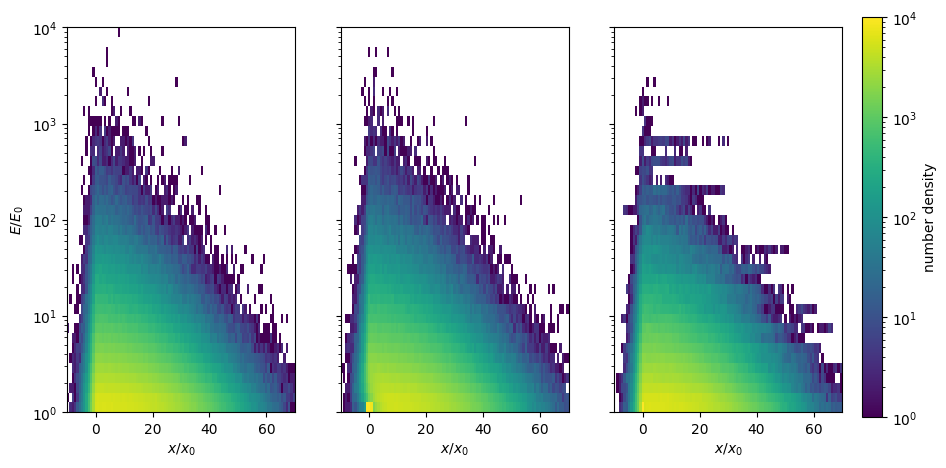}
    \caption{Histogram of the resulting number density $p^2f(p,x)$ with the assumption $E = p/\mathrm{c}$. Left: Continuous injection of $N = 10^6$ candidates at $ x=0$ until $T_{\mathrm{max}} = 100$. Middle: $N_{\mathrm{t}} = 10^5$ candidates are integrated over $N_{\mathrm{obs}} = 10$ snapshots. The resolution in time is not high enough to approximate early times, leading to an overestimation of the source at $X = 0, E = 1$. Right: $N_{\mathrm{t}} = 10^4$ candidates are integrated over $N_{\mathrm{obs}} = 100$ snapshots. }
    \label{fig:timeint-hist}
\end{figure}

\begin{figure}[htbp]
    \centering
    \includegraphics[width=0.8\linewidth]{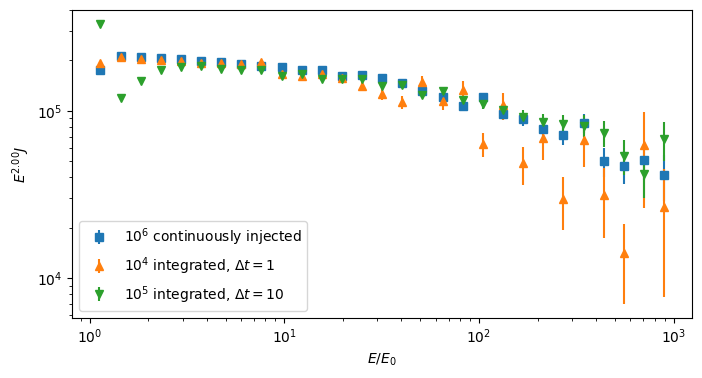}
    \caption{Energy spectra at the shock $x = 0$. All simulations reproduce a $-2$ spectral slope with exponential cut-off. Depending on the number of independent candidates and resolution in the time integration, low or high energies are not represented correctly. The uncertainty grows with decreasing number of independent particles. The uncertainty due to the integration in time is not taken into account. }
    \label{fig:timeint-spectra}
\end{figure}

We compare three simulations: One with continuous injection of $N = 10^6$ independent candidates, one with $N_{\mathrm{t}} = 10^5$ candidates that are integrated with $\Delta T = 10$ over $N_{\mathrm{obs}} = 10$ snapshots, and one with higher resolution in the time integration $\Delta T = 1$, $N_{\mathrm{obs}} = 100$ but less independent candidates $N_\mathrm{t} = 10^4$. The simulation results at $T = 100$ are shown in fig.~\ref{fig:timeint-hist} and the respective spectra at the shock in fig.~\ref{fig:timeint-spectra}. The best result, with lowest oscillations and uncertainties, is obtained by the continuous injection of independent candidates. However, it is computationally the most expensive. With a factor of $10$ less candidates, similar results are obtained in the core of the spectrum. The resolution of the time integration $\Delta T = 10$, however, is not sufficient at early times, when the solution changes quickly. This leads to an overestimation of the source at $x = 0$, $p = 1$. Increasing the time resolution leads to a better approximation around the source and is sufficient to approximate the stationary solution at low energy. Decreasing the number of independent candidates leads to oscillations, visible in the space-energy histogram as horizontal stripes. The number of independent candidates and resolution in time integration have to be chosen carefully.

\section{Validation and Examples}
\label{sec:examples}
This section discusses important parts of the implementation of the solver in more details, including the field line integration (see section \ref{ssec:fieldline}) allowing for diffusion in arbitrary magnetic background fields, an explanation of the uncertainty estimation in more complex simulations (see sections \ref{ssec:cont_injection} and \ref{ssec:splitting}) and several examples on acceleration (see sections \ref{sssec:redsa} and \ref{ssec:MomentumDiffusion}) and pitch angle diffusion (see section \ref{ssec:pitch_angle}). For tests on the correct implementation of the diffusion in homogeneous magnetic fields we refer to section 3.1 in \cite{Merten-etal-2017}. The tests have been repeated with the newly implemented code and no deviation could be found.

\subsection{Field Line Integration}
\label{ssec:fieldline}
As shown in \ref{ssec:sdesolver} the stochastic part of the SDE, namely the diffusion, is solved in the frame of the magnetic field line. This frame is approximated by the local trihedron consisting of the tangential $\textbf{e}_t$, normal $\textbf{e}_n$, and binormal $\textbf{e}_b$ vector of the coherent background field line at each point in space (see \cite{Merten-etal-2017} and references therein). The three vectors are
\begin{align}
    \textbf{e}_t &= \textbf{B} / B \quad , \\
    \textbf{e}_n &= (\textbf{e}_t \cdot \nabla) (k \textbf{e}_t) \quad\text{and}, \\
    \textbf{e}_b &= \textbf{e}_t \times \textbf{e}_n \quad , 
\end{align}
where $k$ is the curvature of the magnetic field line.

In principle, the local trihedron can be calculated and the diffusive step being performed in this frame. However, a problem arises from the stochastic nature of the propagation step. For a given propagation time step $h$ the spatial step, e.g.\ in parallel direction $L = \sqrt{2\kappa_\parallel h} \eta$, can become very large, when a large random number $\eta$ is drawn and the local trihedron can change significantly during one propagation step. So naively propagating only in the tangential direction could lead to very large deviations from the actual field line for a pure parallel diffusion scenario.

The propagation in parallel direction is therefore performed by a field line integration:
\begin{align}
    \Delta \mathbf{x_\parallel} = \int_0^L \textbf{e}_t(s) \,\mathrm{d}s \quad, \label{eq:fieldline}
\end{align}

where $s=0$ parameterizes the start point of the field line integration $\textbf{x}_n$. The integration is performed with an adaptive 4-th order Runge-Kutta algorithm using a fifth order approximation to estimate the local error and adjust the step length. The propagation distance $L$ can be split in several subsets if necessary (see \cite{Merten-etal-2017} for more details). 

When the parallel step is performed the perpendicular propagation is calculated in the plane that is defined by $\Delta \mathbf{x_\parallel})$. Up to know it is assumed that the two perpendicular directions are degenerated ($\kappa_{\perp, 1} = \kappa_{\perp, 2}$), allowing to randomly choose the normal and binormal vector. This approximation is usually valid for Galactic transport but has to be revised for, e.g., the propagation of low energetic particles in the Solar magnetic field. 

\begin{figure}[htbp]
    \includegraphics[trim={1.5cm 0 2cm 1.1cm}, clip, width=0.5\linewidth]{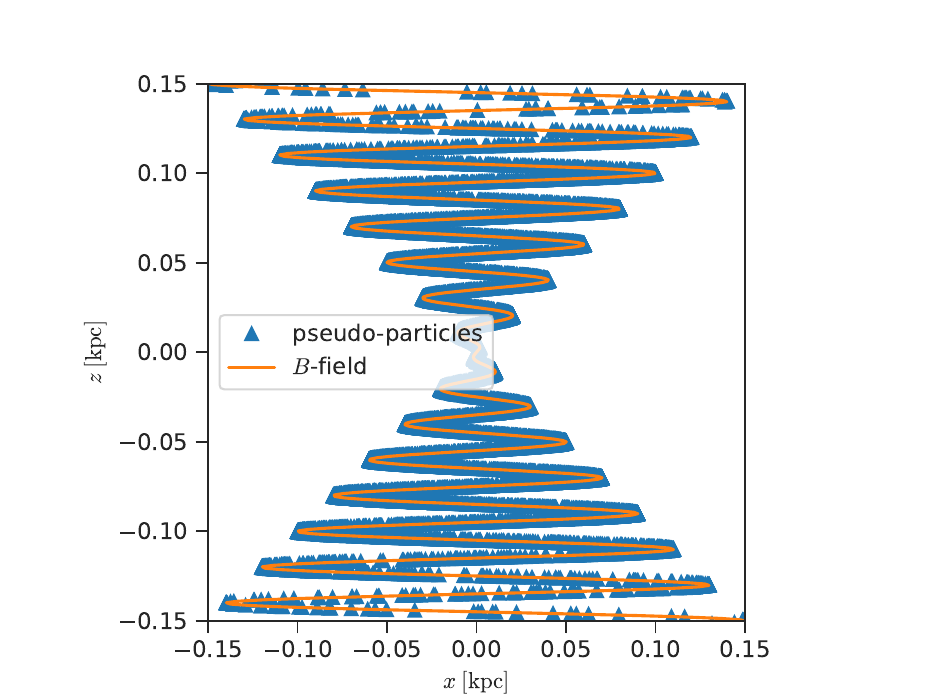}
    \includegraphics[trim={1.5cm 0 2cm 1.1cm}, clip, width=0.5\linewidth]{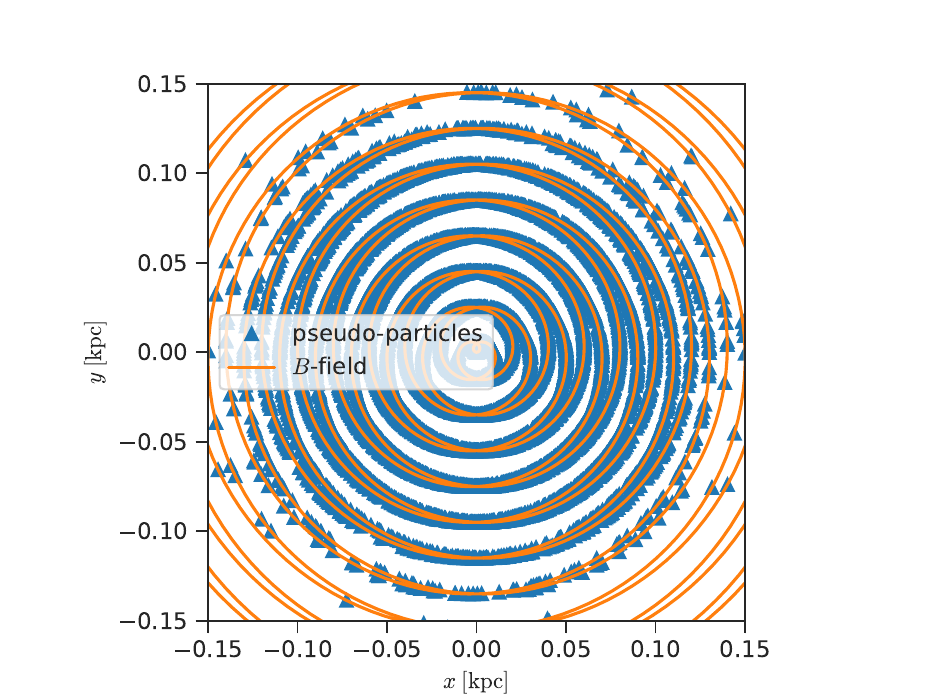}
    \caption{Pseudo-particle end positions for pure parallel diffusion (blue, triangles) in comparison to the analytical magnetic field line (orange line). Projection into the $x$-$z$-plane of the three dimensional simulation of a spiral with changing curvature radius is shown.}
    \label{fig:fieldlineintegration}
\end{figure}

An example of pure parallel diffusion can be seen in figure \ref{fig:fieldlineintegration}. For this validation pseudo-particles are injected at the origin and then propagated with pure parallel diffusion ($\kappa_\perp=0$) and without advection. It can be clearly seen, that the pseudo-particles stick to their original field line given by:
\begin{align}
    \mathbf{r}_\mathrm{spiral} = z (\cos(2\pi z/s)\,\mathbf{e}_x + \sin(2\pi z/s)\,\mathbf{e}_y + \mathbf{e}_z) \quad ,
\end{align}
where $s$ is a parameter, defining how strongly the spiral winding around the $z$-axis. The interested reader is referred to the appendix \ref{apdx:fli} for an analysis of the uncertainty connected to the field line integration. 

\subsubsection{Focusing length}
The focusing length as defined in \cite{vandenBerg-etal-2020} can be derived approximately by:
\begin{align}
    L = - B  \left( \frac{\partial B}{\partial s} \right)^{-1} \approx \frac{B(x_{n+1}) - B(x_n)}{\Delta s} \quad ,
\end{align}
where $\Delta s = \eta_t \sqrt{2\kappa_\parallel h}$ is the tangential step length along the field line direction $\textbf{e}_t=\textbf{B}/B$. This value can be calculated in each propagation time step allowing for a spatial dependence of the focusing length. The discussion of a spatially varying focusing length is beyond the scope of this paper but refer to section \ref{ssec:pitch_angle} for an in depth analysis of the relevance of constant focusing.

\subsection{Diffusive shock acceleration}
\label{sec:DSA}

In the ensemble-averaged description of DSA, energy gain at the shock comes from the adiabatic compression of the background flow. Depending on the interplay of diffusion and advection, the expected spectra at the shock are obtained.\footnote{This is equivalent to the description of the escape probability and compression ratio, which determine the spectral slope in case of a 1D stationary planar shock when individual particles are considered.} As discussed in sec.~\ref{sssec:adaptive_dsa}, this leads to constraints on the time step, since pseudo-particles have to encounter the shock region during the simulation, but on the other hand, the diffusive length scale still has to be significantly larger than the shock width. This is discussed in detail in  \cite{Kruells-Achterberg-1994, Achterberg-Schure-2010, Aerdker-etal-2024}.

\subsubsection{Re-acceleration at a Galactic Wind Termination Shock}
\label{sssec:redsa}

In this scenario CRs are re-accelerated at a spherical Galactic wind termination shock (GWTS) at a distance of $250\,\mathrm{kpc}$ \citep{Merten-etal-2018}. CRs pre-accelerated in the Galactic disk are advected outwards until they encounter the GWTS. Given the large shock radius and considering magnetic field amplification close to the shock, particles with rigidity up to $10^{17}\,\mathrm{V}$ can still be confined (see e.g.~\cite{Bustard-etal-2017}). Most of the particles will be advected outwards, but a fraction is able to propagate back to the Galaxy and contributes to the spectrum between the knee and the ankle \citep{Merten-etal-2018}.

\begin{figure}[htbp]
    \centering
    \includegraphics[width=0.49\textwidth]{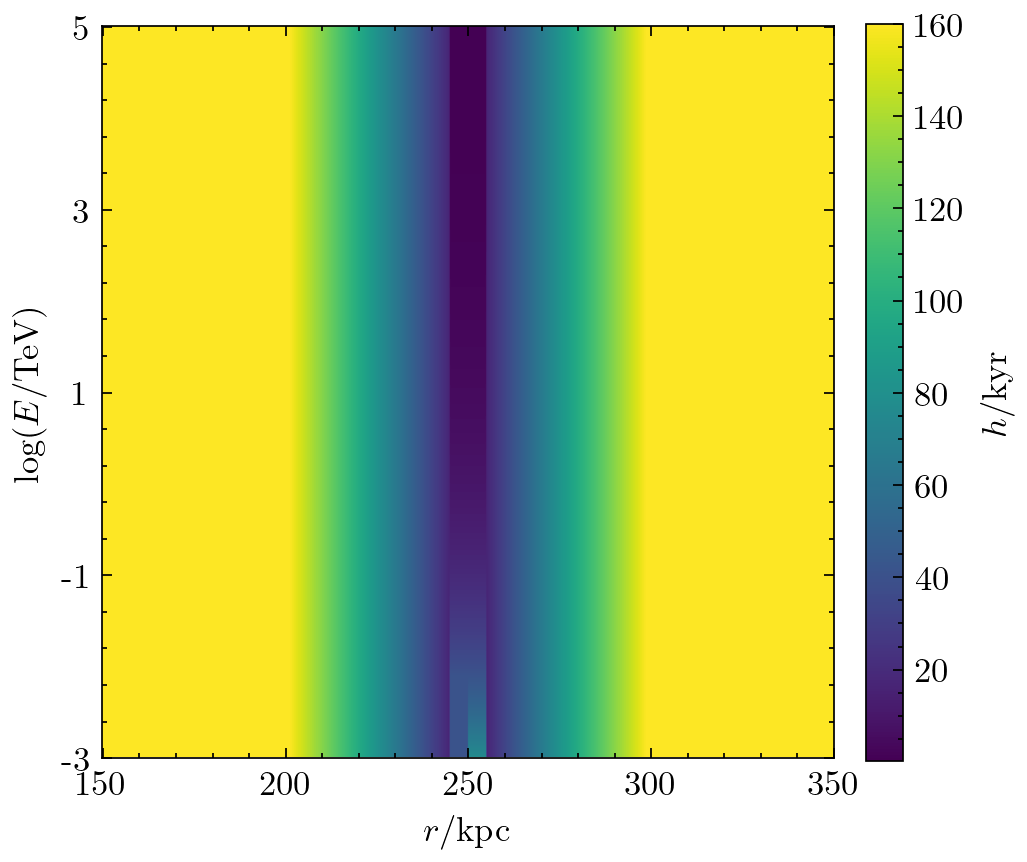}
    \includegraphics[width = 0.49\textwidth]{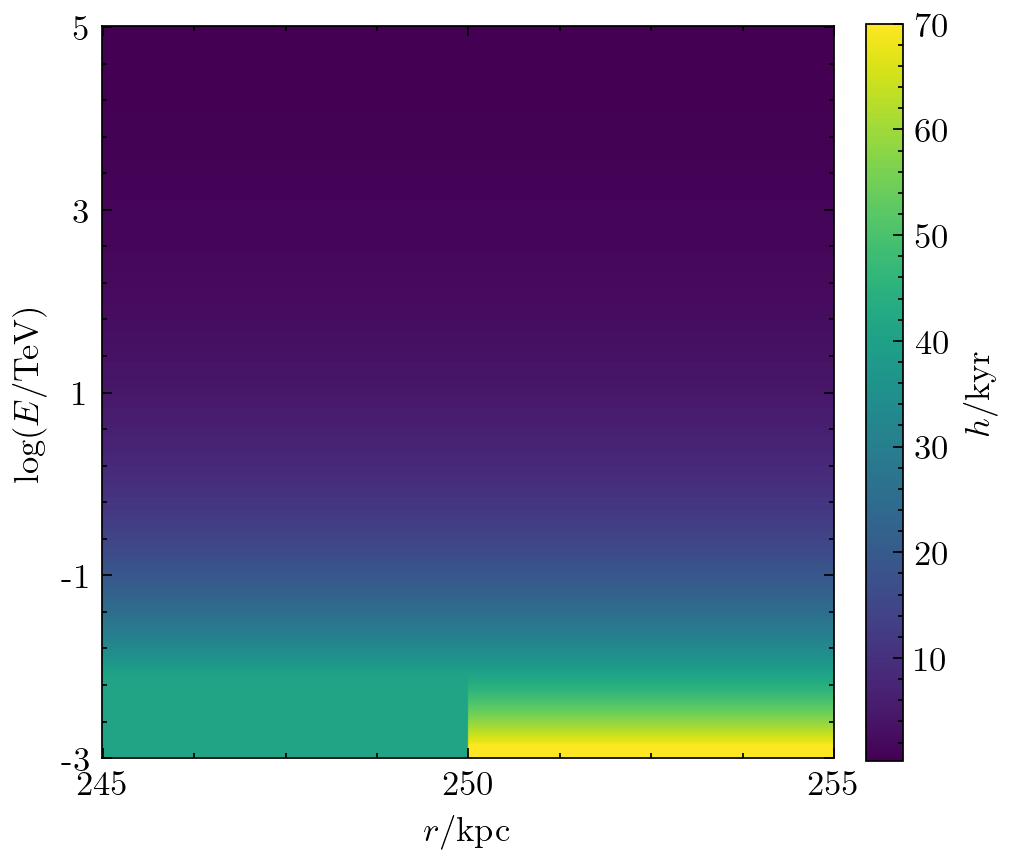}
    \caption{ Adaptive integration step $h$, depending on the pseudo-particles energy and distance to the shock. Left: Close to the shock $r = [245, 255]\,\mathrm{kpc}$ the inequality \ref{eq:Achterberg} determines the step. Outside that region the maximal step is used but limited to the distance to the shock. Right: Close view at the shock region. With increasing energy, the limiting factor of the inequality changes from the advective step (for low energies) to the diffusive step (for high energies). The advection speed drops at the shock, allowing for larger time steps. }
    \label{fig:adaptive-step}
\end{figure}

For a simulation that includes both, transport out and back to the Galaxy, as well as acceleration at the shock, the integration time step needs to change from $\approx 100\,\mathrm{pc}$ to $ > 1\,\mathrm{pc}$ in the shock region. Figure \ref{fig:adaptive-step} shows the chosen time step depending on the pseudo-particles energy and distance to the spherical shock. The considered magnetic field is radial to show the effect of the adaptive step with respect to the constraints given by DSA only. In curved magnetic fields, the integration step also depends on the local curvature of the magnetic field lines, see \cite{Merten-etal-2017} for details. 

In this example several methods to increase statistics/lower computation time are combined: A flat energy spectrum is injected and re-weighted in the analysis for better statistics at high (injected) energies. The \texttt{CandidateSplitting} module is used to increase statistics at even higher energies when candidates gain energy at the shock. An adaptive integration step is used, to meet the constraints discussed in sec.~\ref{sssec:adaptive_dsa} close to the shock and to have a sufficiently large integration step far away from the shock. The results are integrated over time (see sec.~\ref{ssec:cont_injection}) saving computation time as only a fraction of pseudo-particles have to be injected compared to continuous injection.

\begin{figure}[htbp]
    \centering
    \includegraphics[width=0.8\textwidth]{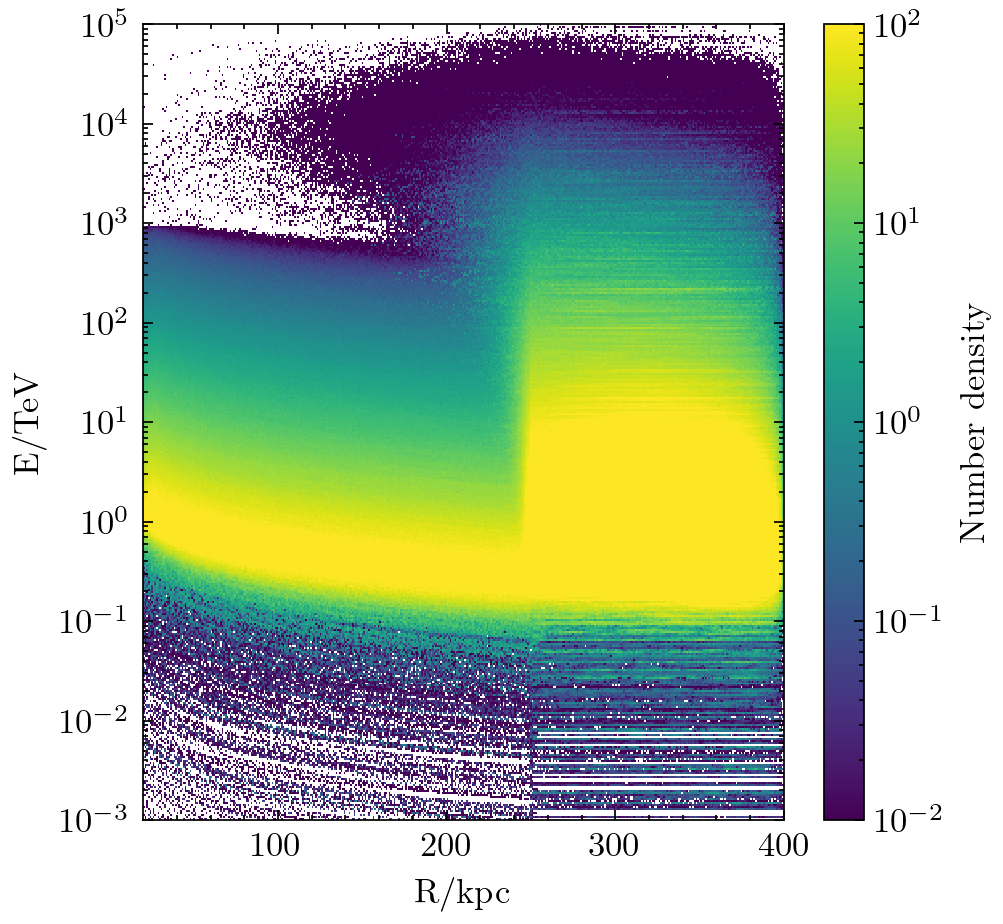}
    \caption{Number density $n=fp^2$ after $4\,\mathrm{Gyr}$. Pseudo-particles are injected at $r = 20\,\mathrm{kpc}$ with energies ranging from TeV to PeV. Pseudo-particles are cooled due to the expanding wind and are re-accelerated when they encounter the shock at $250\,\mathrm{kpc}$.  }
    \label{fig:GWTS}
\end{figure}

The resulting number density after $4\,\mathrm{Gyr}$ is shown in fig.~\ref{fig:GWTS}. Pseudo-particles are injected at a sphere of $r = 20\,\mathrm{kpc}$ with energies ranging from TeV to PeV. Here, a spectral slope of $-2$ is assumed for the injected spectrum but can be re-weighted to any other power-law. Diffusion is energy-dependent $\kappa(E) = 5 \times 10^{24}\,\mathrm{m^2/s}\, (E/\mathrm{GeV})^{0.3}$. Particles are advected out and cooled in the expanding wind. High-energy particles propagate faster due to the higher diffusion coefficient and experience less cooling. Once the particles encounter the shock, they are re-accelerated and slowed down in the downstream region, away from the Galaxy. The upstream distribution of re-accelerated particles is well visible at $R < 250\,\mathrm{kpc}$ and $E > 10^3\,E_0$. It is more likely for high-energy particles to propagate back to the Galaxy.

\subsection{Superdiffusion}
\label{ssec:Superdiffusion}

Non-Gaussian particle transport, characterized by a nonlinear dependence of the mean squared displacement on time, can be modeled by fractional transport equations, like equ.~\ref{eq:space-fractional}. Here, we focus on space-fractional transport describing superdiffusion,  $\langle \Delta x^2\rangle \propto t^{\zeta}$, $\zeta > 1$. 

\begin{figure}
    \centering
    \includegraphics[width=0.5\linewidth]{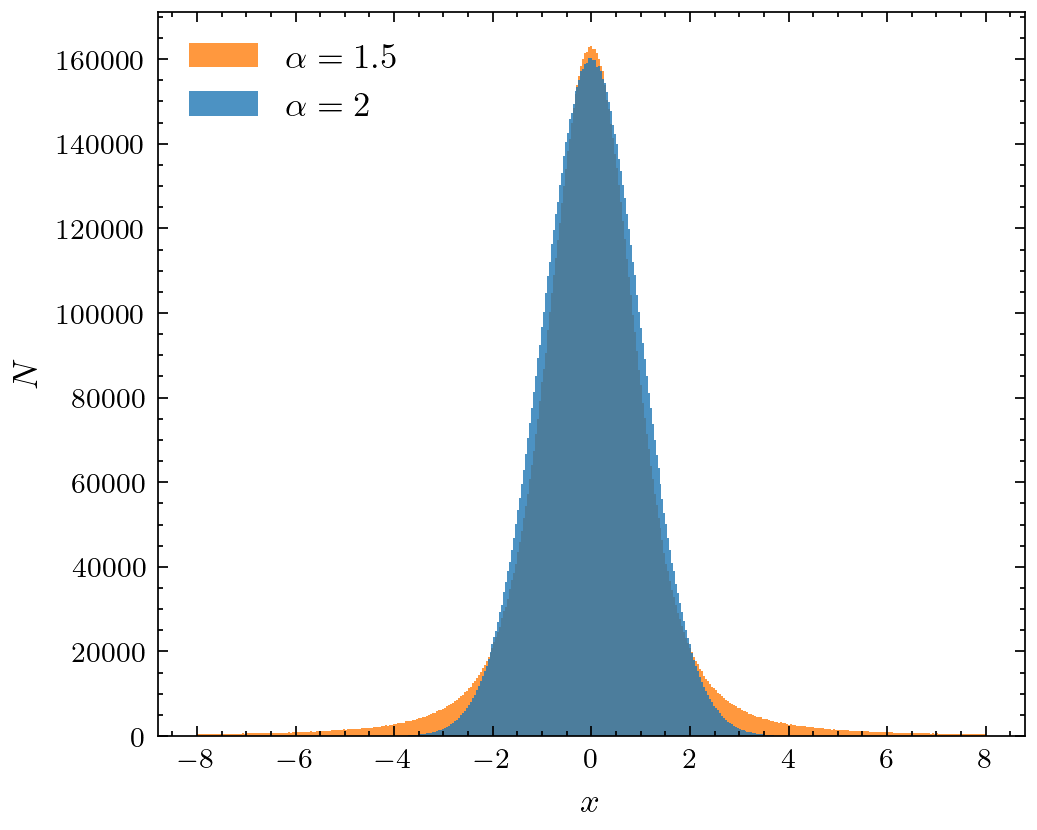}
    \caption{Histogram of random numbers that are drawn from an alpha-stable L\'evy distribution, for $\alpha = 1.5$ and $\alpha = 2$ (equivalent to normal distribution). }
    \label{fig:Levydistr}
\end{figure}

\begin{figure}
    \centering
    \includegraphics[width=0.49\linewidth]{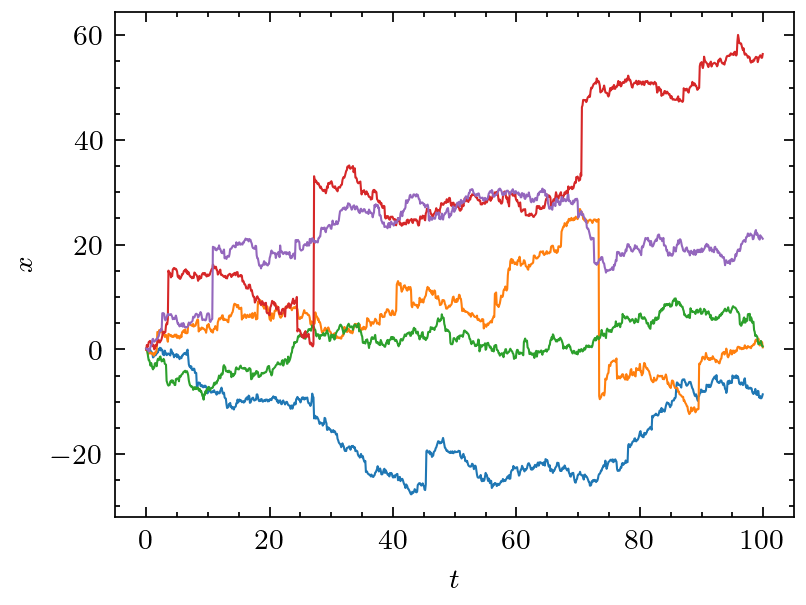}
     \includegraphics[width=0.49\linewidth]{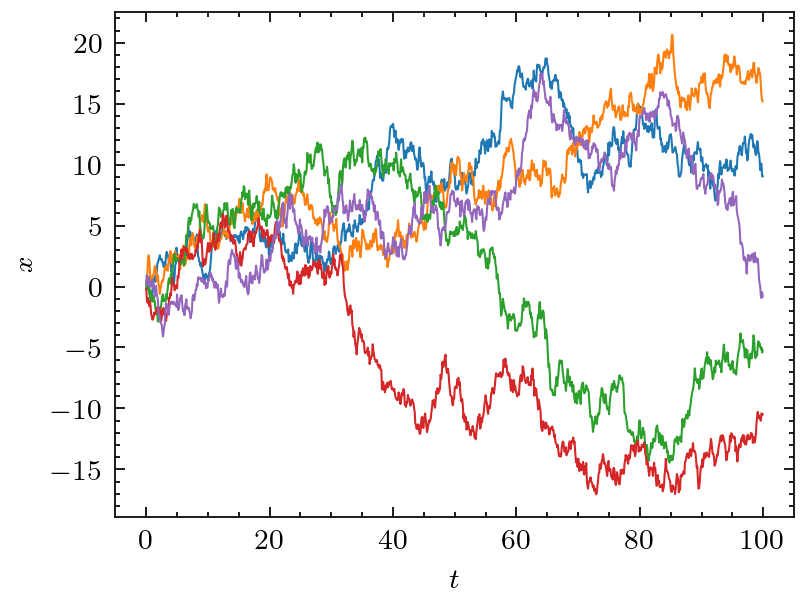}
    \caption{Five example pseudo-particle trajectories with L\'evy flights (left) and normal diffusion (right). The spatial displacement $x$ e.g.~ along a magnetic field against the normalized time $t$ is shown. The (anomalous) diffusion coefficient is $\kappa_{\alpha} = 1$.}
    \label{fig:Levyflights}
\end{figure}

In the SDE approach, each simulation step the random number $\eta$ is drawn from an alpha-stable L\'evy distribution, which has enhanced tails compared to the normal distribution (see fig.~\ref{fig:Levydistr}). Occasionally, a large number is drawn from the tails and the pseudo-particle experiences a L\'evy flight\footnote{Such L\'evy flights can lead to superluminal speed of the \emph{pseudo-particle}. However, this is also already true for Gaussian diffusion as discussed in sec.~\ref{ssec:pitch_angle}}. Figure \ref{fig:Levyflights} shows 5 pseudo-particle trajectories with Gaussian diffusion and L\'evy flights, where $\alpha = 1.7$. The fractional dimension $\alpha = 2/\zeta$ of the transport equation characterizes the frequency and length of such jumps. For $\alpha = 2$, the normal distribution is recovered.

Superdiffusive transport leads to power-law distributions in space, e.g.~at the upstream sides of shocks. In simple cases, e.g.~diffusion-advection equations, the resulting distribution function can also be approximated by a Fourier series, which is used to validate the algorithm. For details we refer to \cite{Effenberger-etal-2024, Aerdker-etal-2024b}, where the transport and acceleration of particles at shocks are investigated.

\subsection{Momentum Diffusion --- Second Order Fermi Acceleration}
\label{ssec:MomentumDiffusion}
Momentum diffusion is the ensemble-averaged description of second order Fermi acceleration. In contrast to diffusive shock acceleration (DSA) (see, e.g., \citep{Aerdker-etal-2024}), individual particles of the ensemble can loose energy, making the process in realistic physical situations slower. Nevertheless, this form of stochastic acceleration can play an important role, e.g., in compact sources or in the Galaxy (see e.g.\ \cite{Drury-Strong-2017, Cadillo-2019}). 

Assuming a stationary transport equation ($\partial_t n=0$) that neglects all terms but momentum diffusion the equation reads
\begin{align}
    0 = \frac{\partial}{\partial p} \left[p^2 D_0p^{\alpha_p} \frac{\partial}{\partial p} \left( \frac{n}{p^2} \right) \right] \quad .\label{eq:statMomentumDiff}
\end{align}
Here, it is assumed that the momentum diffusion scalar is described by a single power law $D_{pp}=D_0p^{\alpha_p}$ in momentum $p$ and the particle number density is given by $n$. By integrating this differential equation one can derive the momentum dependence of the stationary particle number density $n \propto p^{1-\alpha_p}$. Therefore, harder\footnote{Harder spectra refer to larger spectral indices, leading to more high energetic particles.} spectra are produced by momentum diffusion than by DSA. 

To validate the simulation code we modeled the particle number density coming from pure momentum diffusion; neglecting all other terms of the transport equation. In doing so, the momentum diffusion scalar was a) constant to $D_{pp}=1\,\mathrm{N}^2\,\mathrm{s}$ and b) had a small momentum dependence $D_{pp}\propto p^{1/3}$

\begin{figure}[htbp]
    \begin{minipage}{.49\textwidth}
            \includegraphics[width=\textwidth]{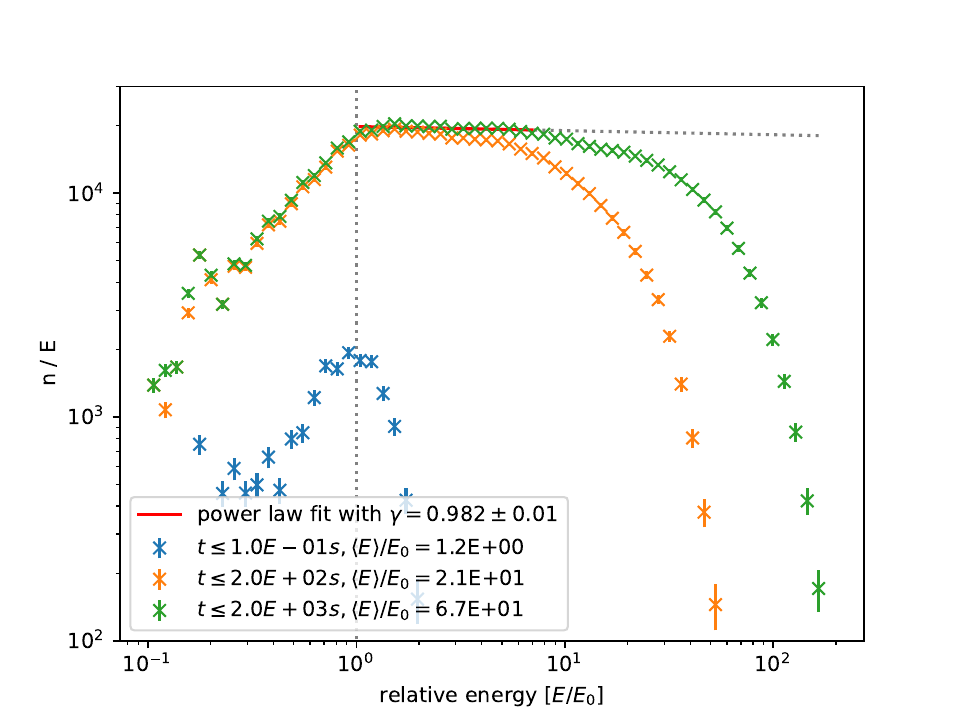}
    \end{minipage}%
    \begin{minipage}{.49\textwidth}
            \includegraphics[width=\textwidth]{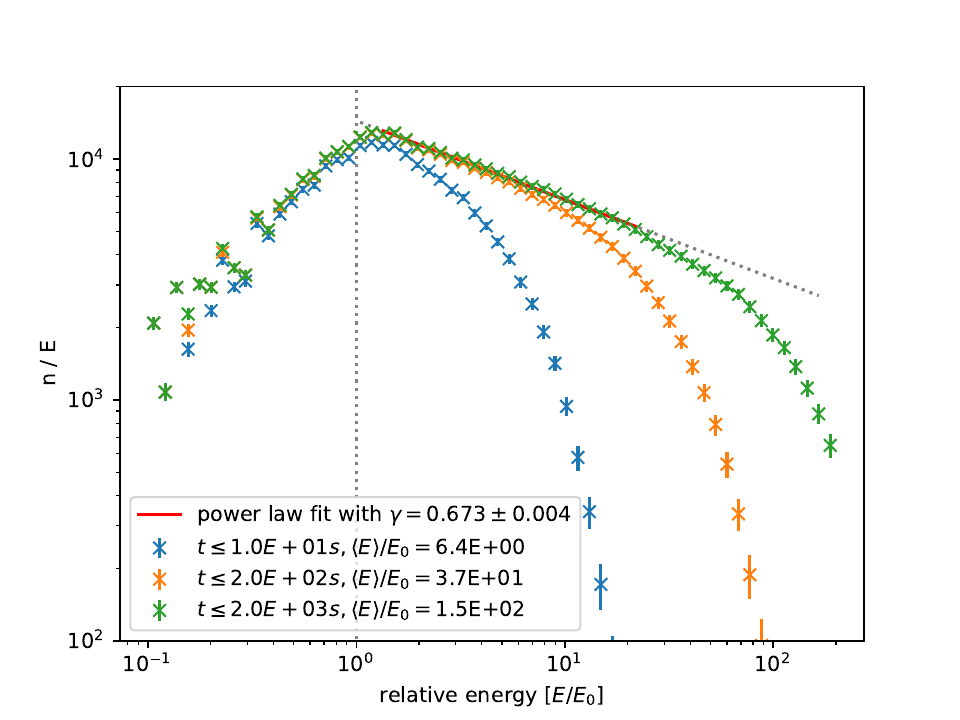}
    \end{minipage}
    \caption{Cosmic-ray energy distribution assuming pure momentum diffusion is shown for different times. Here, a continuous injection of CRs with energy $E_0$ is modeled. At energies above the injection scale the expected power law behavior (left constant diffusion coefficient and right $D_{pp}\propto p^{1/3}$) is visible and fitted (red dashed line). The cut-off due to the finite simulation time is visible at higher energies. For energies below the injection scale a very soft spectrum is expected.}
    \label{fig:MomDif_Const_Mono}
\end{figure}

Figure \ref{fig:MomDif_Const_Mono} shows the time evolution assuming a continuous injection of particles. This approximates the solution of equ.~\ref{eq:statMomentumDiff} since CRPropa has no explicit solver for stationary equations (see sec.~\ref{ssec:cont_injection}). In this run 1000 pseudo-particles have been injected and afterwards recorded $N_\mathrm{obs}$-times. During post-processing the energy spectrum is derived for each of the snapshots and afterwards summed up. The interested reader is referred to \citep{Merten-etal-2018, Aerdker-etal-2024} for more information on this technique. The expected power-law behavior is reached between the injection energy $E_0$ and a time dependent cut-off energy, which comes from the finite simulation time. At energies below $E_0$ the spectrum is much softer.

\subsection{Pitch Angle Diffusion}
\label{ssec:pitch_angle}

\begin{figure}[htbp]
    \begin{minipage}{.49\textwidth}
            \includegraphics[width=\textwidth]{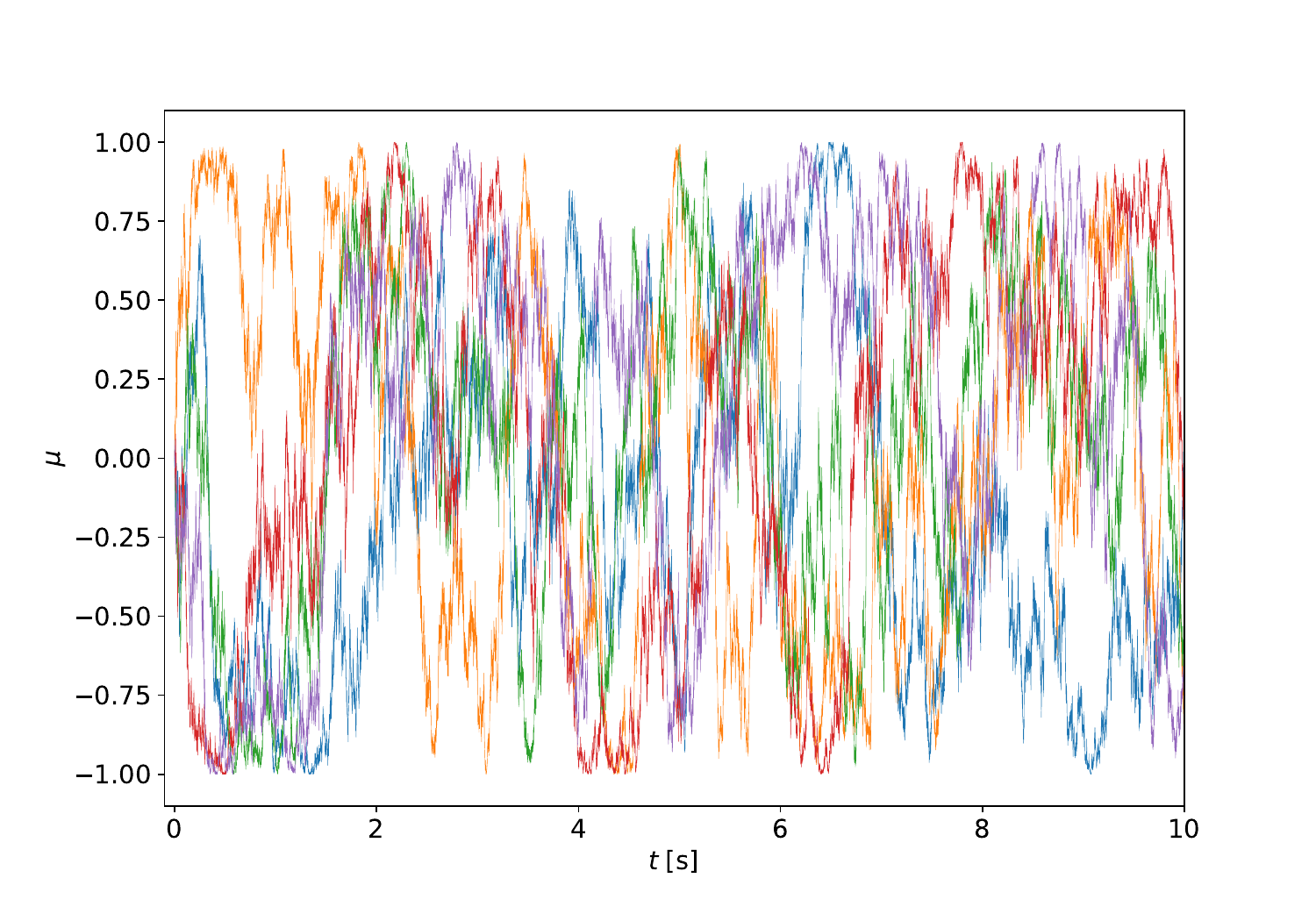}
    \end{minipage}%
    \begin{minipage}{.49\textwidth}
            \includegraphics[width=\textwidth]{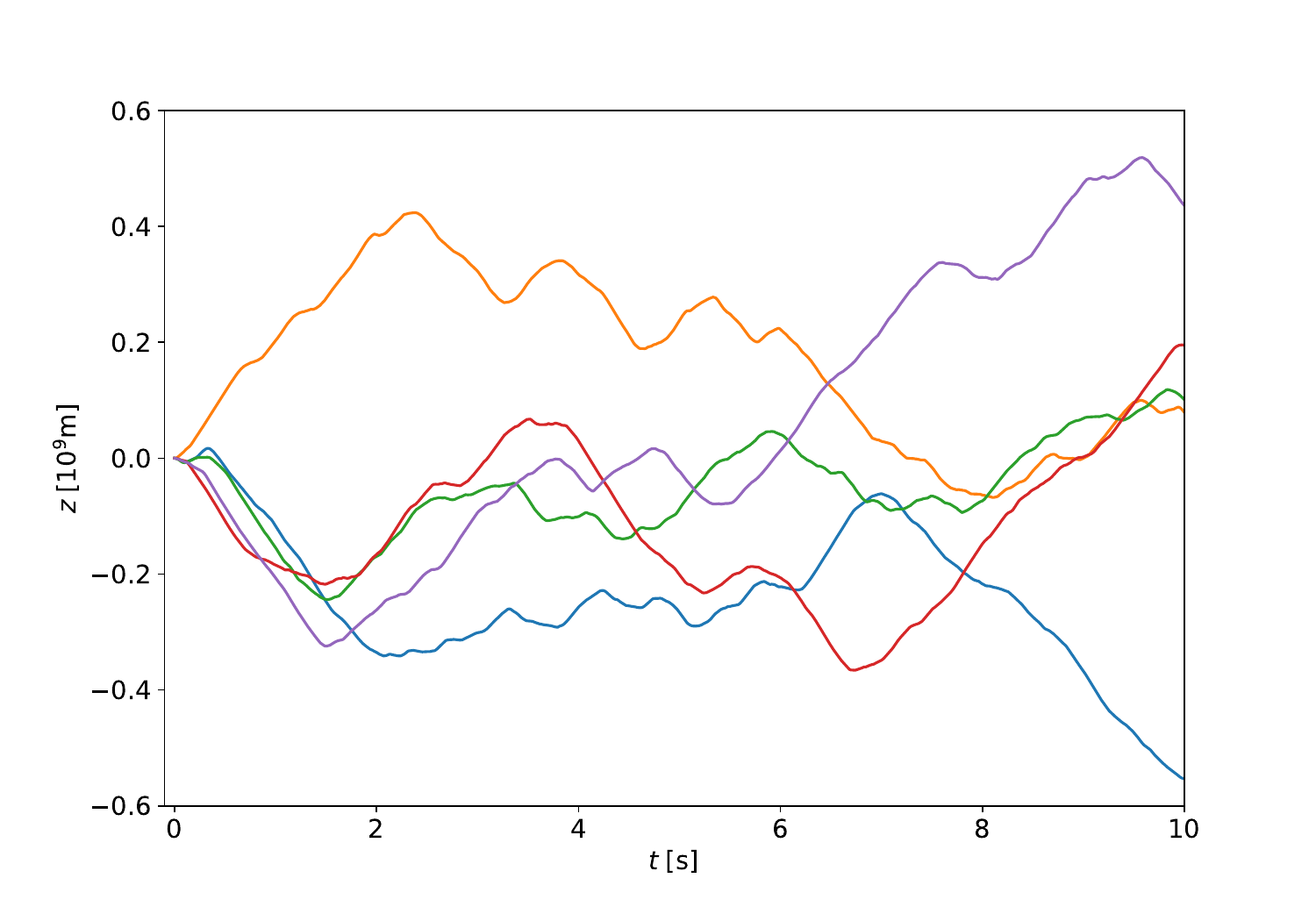}
    \end{minipage}
    \caption{Left figure shows five example pitch angle trajectories. The stochastic nature for an isotropic diffusion coefficient is clearly visible. The right figure shows the corresponding trajectories along the magnetic field line, which show a less pronounced stochastic behaviour. This is due to the fact, that small changes in pitch angle usually do not change the propagation direction of the pseudo-particle.}
    \label{fig:pitchangle_base}
\end{figure}

To illustrate that pitch angle diffusion leads to a correlated random walk along the magnetic background field, we model isotropic pitch angel diffusion with $D_{\mu\mu}^\mathrm{iso} = D_0(1-\mu^2)$. Figure \ref{fig:pitchangle_base} shows five example trajectories in pitch angle $\mu(t)$ and the corresponding position along the field line $z(t)$ --- as the magnetic field was chosen to be $\mathbf{B}=B_0\mathbf{e}_z$. The reflective boundary conditions $(-1, 1)$ are clearly visible in the left part of the figure. Furthermore, the diffusive behavior of the pitch angle is clearly visible. One might divine that the diffusion coefficient $D_{\mu\mu}^{iso}$ is smaller at the boundaries ($\mu = \pm 1$). The right figure shows much less chaotic trajectories in real space, as the small changes in pitch angle lead to a somewhat correlated random walk along the field line. 

Averaging the focused transport equation \ref{eq:focused_transport} over the pitch-angle $\mu$ allows to directly compare the simulation with pure spatial diffusion. The asymptotic behavior of the spatial distribution function is given by the spatial diffusion coefficient (see e.g.~\cite{Shalchi-2009}):
\begin{align}
    \kappa_\parallel &= \frac{v^2}{8} \int_{-1}^1 \frac{(1-\mu^2)^2}{D_{\mu\mu}} \notag \\ 
    &= \frac{v^2}{6D_0} \quad . \label{eq:pitch2spatial}
\end{align}

\begin{figure}[htbp]
    \centering
    \includegraphics[width=\textwidth]{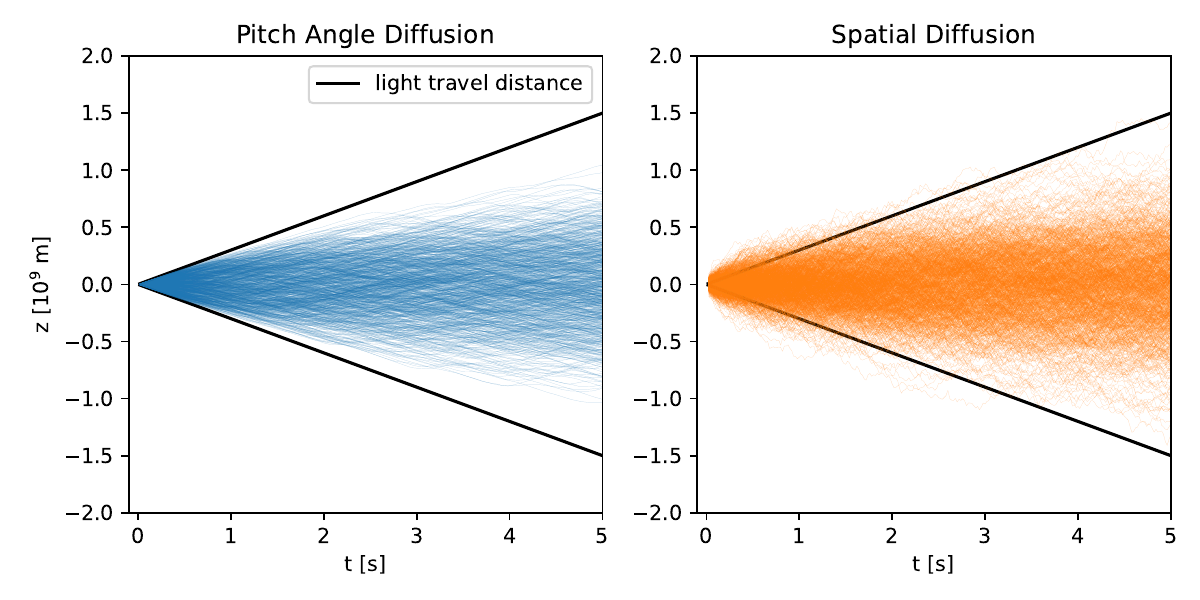}
    \caption{Left figure shows 1000 example trajectories of pseudo-particles modeled with isotropic pitch angle diffusion and isotropic injection. Right figure shows the same number of pseudo-particles trajectories modeled with spatial diffusion parallel to the magnetic background field. The spatial diffusion coefficient was chosen according to equ.~\ref{eq:pitch2spatial}. It can be noted that pitch angle trajectories (left) are always slower than the speed of light (black solid lines), while this is not true for spatial diffusion.}
    \label{fig:comparison_pitch_spatial}
\end{figure}

Figure \ref{fig:comparison_pitch_spatial} shows this comparison for $D_0=1\,\mathrm{N}^{2}\,\mathrm{s}$. Here, 1000 example trajectories are shown for an isotropic injection --- $\mu$ uniformly distributed between -1 and 1 --- at a point source. For pure spatial diffusion (right plot) it can be seen that the pseudo-particles move faster than the speed of light; a common problem of spatial diffusion models. This super-luminal motion of phase space elements is not part of the pitch-angle diffusion model, since here the speed along the field line is bound by the real particle speed $v$. This difference between the models is clearly visible in the figure, where the pitch angle trajectories do not cross the lines defined by the speed of the particles ($|z_\mathrm{light}(t)| = c_0 t$, black solid lines). This effect is more pronounced at early times.  

\begin{figure}[htbp]
    \centering
    \includegraphics[width=\textwidth]{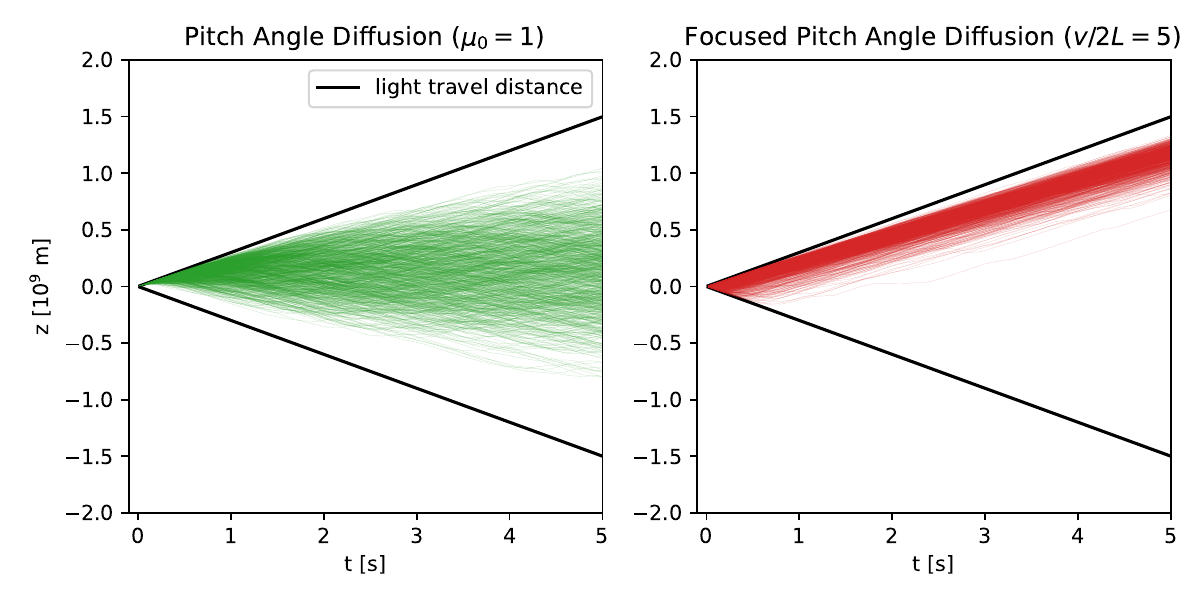}
    \caption{Similar to fig. \ref{fig:comparison_pitch_spatial} the left plot shows 1000 trajectories of isotropic pitch angle diffusion with an anisotropic injection. It can be seen that after a short time an isotropic pitch angle distribution is reached and the asymptotic behavior is similar to the isotropic injection scenario. In contrast the right figure displays 1000 trajectories of a constant focusing model. Here, it can be seen that an isotropic injection distribution is quickly focused in one preferred direction leading to an average drift along the field line. Also the spread of the trajectories is smaller.}
    \label{fig:comparison_focus}
\end{figure}

When changing the injection to a fully anisotropic distribution $\mu_0=\mu(t=0)=1$ the results are only affected at early times (see fig.\ \ref{fig:comparison_focus}, left plot). The phase space elements move at the beginning as a narrow strahl along the magnetic field line. After a short time ($\sim 2\,\mathrm{s}$) the pitch angle diffusion leads to an isotropic distribution and almost no differences to the isotropic emission scenario are found. In contrast to that, the focusing has a larger effect on the asymptotic behavior. For a spatially constant focusing with $v/(2L)=5$ it can be clearly seen (right panel of fig.~\ref{fig:comparison_focus}) that the spread of the particle distribution gets smaller, hence the name focusing term. The particle distribution collectively drifts with an average velocity along the magnetic field line. Note, that this motion is not given by the gradient-B drift which would be perpendicular to the magnetic field direction. The speed of this drift along the field line $v_\mathrm{drift}$ can be calculated by means of a fixed point analysis of the focused transport equation \ref{ssec:fixpoints}.

\begin{figure}[htbp]
    \centering
    \includegraphics[width=.8\textwidth]{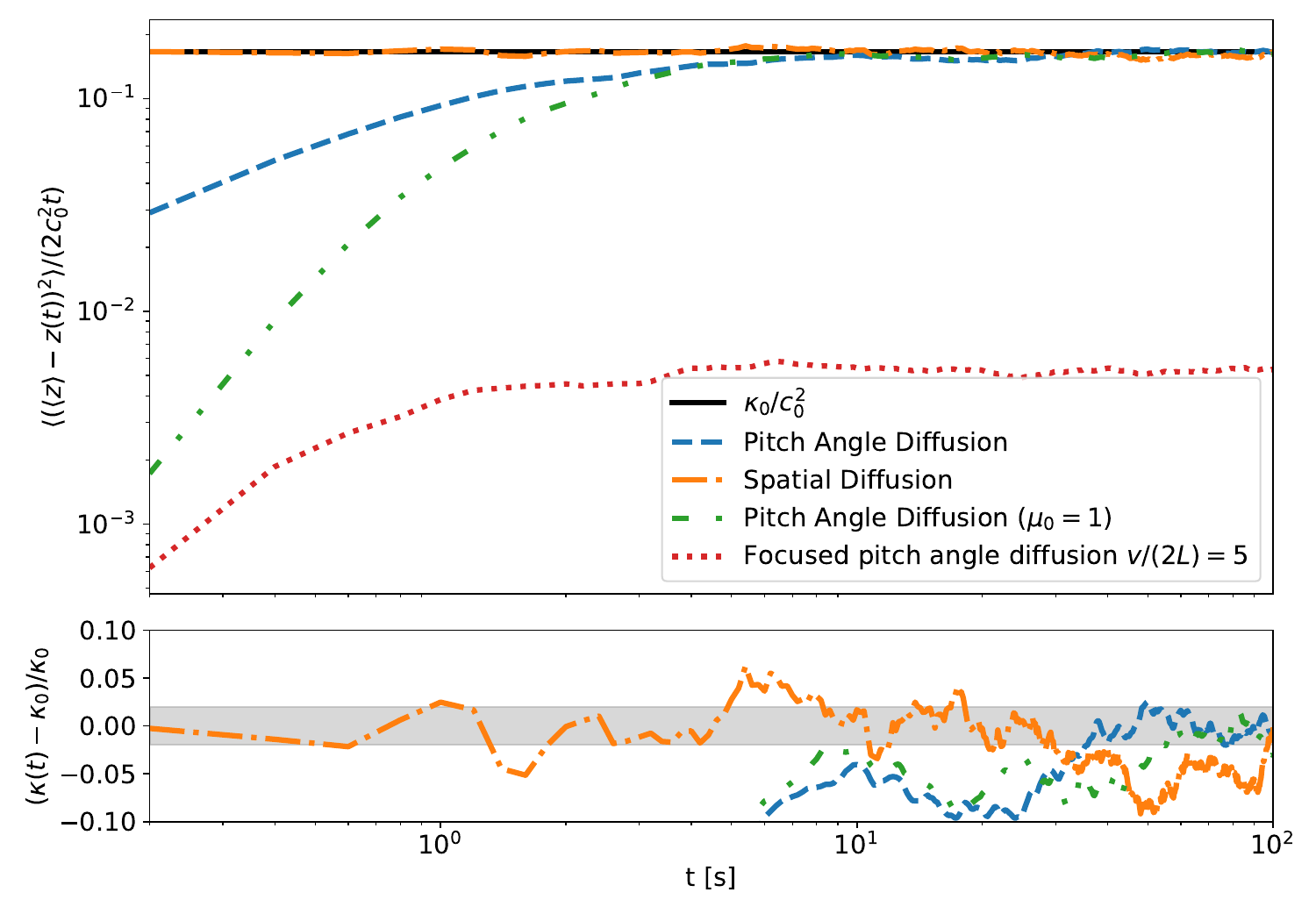}
    \caption{Running diffusion coefficient (upper panel) and ratio plot (lower panel) comparing the four different models of particle transport along a field line --- isotropic pitch angle diffusion (blue dashed), pitch angle averaged spatial diffusion (orange dash-dotted), pitch angle diffusion with anisotropic injection (green dash-dot-dotted) and focused pitch angle diffusion (red dotted). }
    \label{fig:enter-label}
\end{figure}

\begin{figure}[htbp]
    \centering
    \includegraphics[width=.8\textwidth]{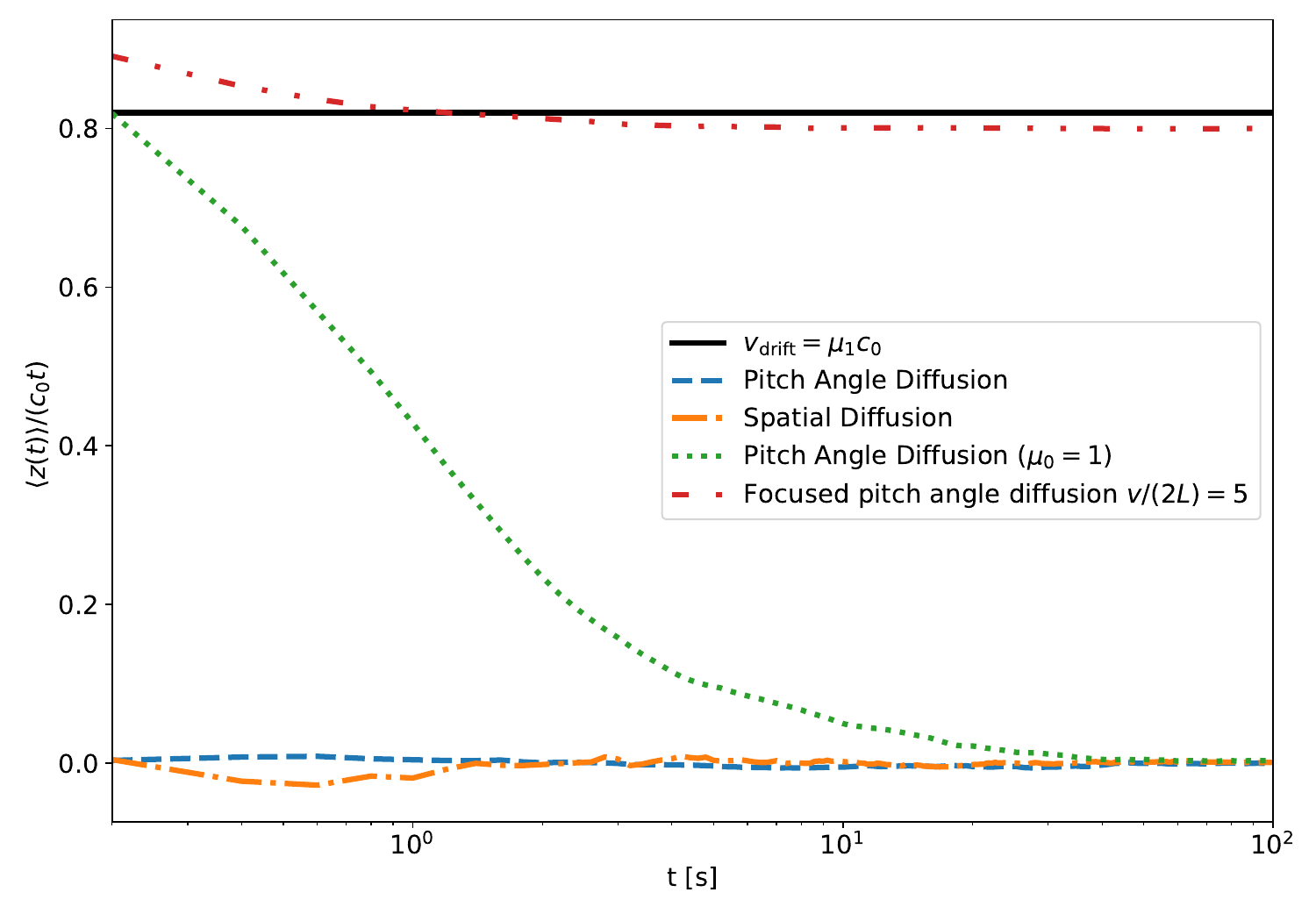}
    \caption{Normalized mean propagation speed along the field line, comparing the four different models of particle transport along a field line --- isotropic pitch angle diffusion (blue dashed), pitch angle averaged spatial diffusion (orange dash-dotted), pitch angle diffusion with anisotropic injection (green dash-dot-dotted) and focused pitch angle diffusion (red dotted). Speed calculated with with the fixed point method is shown in black.}
    \label{fig:meanpropagation}
\end{figure}

\subsubsection{Fixed Points}
\label{ssec:fixpoints}

The stochastic differential equation for the pitch-angle reads
\begin{align}
\label{eq:dmudt}
    \mathrm{d}\mu = \frac{v}{2 L} \left(1 - \mu^2\right) \mathrm{d}t+ \frac{\partial D_{\mu\mu}}{\partial \mu}\mathrm{d}t + \sqrt{2 D_{\mu\mu}} \mathrm{d}W_{t} \quad,
\end{align}
where $D_{\mu\mu} = D_0 (1-\mu^2)$ is non-linear. 

When the focusing term is neglected $(L \rightarrow \infty)$, equ.~\ref{eq:dmudt} is a linear SDE and has a stable fixed point at $\mu = 0$. The pitch-angle is pulled back to $\mu = 0$, the strength of this restoring force depends on the value of $D_0$ and $\mu$ itself. This way, the dynamical system at hand prevents the pitch-angle to diffuse to arbitrary values just like the reflective boundaries at $\mu = \pm 1$\footnote{One may speculate if the boundaries would be necessary given a sufficiently small integration step $h \rightarrow 0$.}. Note, that for individual pseudo-particles, the stochastic motion governed by $\sqrt{2 D_{\mu\mu}} \mathrm{d}W_{t}$ would kick the pitch-angle out of the stable fixed point again even if it would be reached at some point. On average, however, the stochastic term cancels out and we expect the mean pitch-angle to be zero. 

\begin{figure}[htbp]
    \centering
    \includegraphics[width=.6\textwidth]{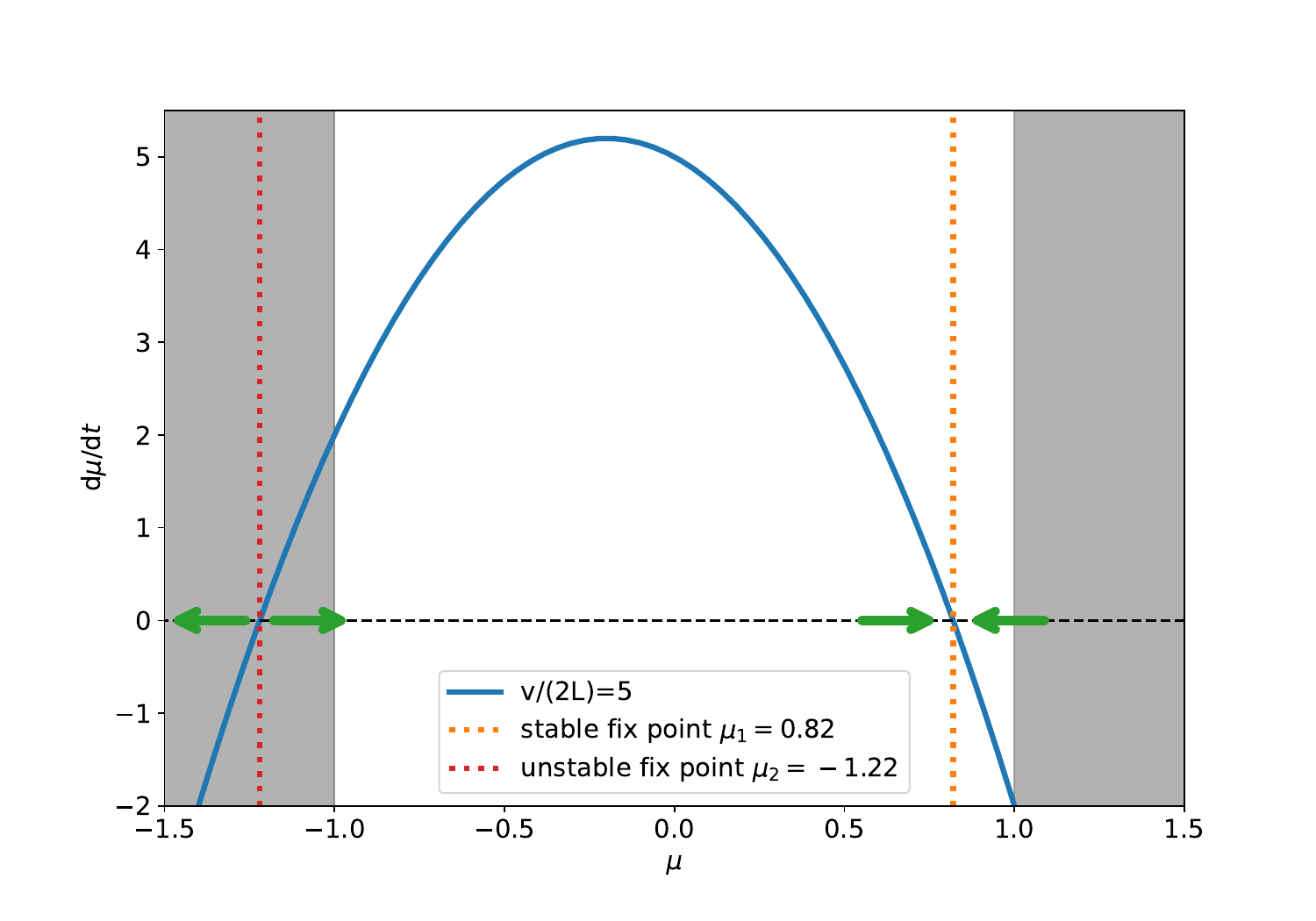}
    \caption{Example fixed point analysis for $v/(2L) = 5$ and $D_0=1$. This model has two fixed points where the stable one (orange dashed line) is within the allowed range of pitch angle values. The unstable fixed point (red dashed line) is smaller than $\mu<-1$ and cannot be reached.}
    \label{fig:fixpoints}
\end{figure}

The dynamics become more interesting when the focusing term is included. The now non-linear SDE has two fixed points: the previous stable one is pushed to positive values of $\mu$ and accompanied by an unstable fixed point at $\mu < 0$. The unstable fixed point can lie outside the boundaries depending on the systems parameters, while the stable fixed point is between $\mu = 0$ and $\mu = 1$. Figure \ref{fig:fixpoints} visualizes the fixed point of the dynamical system. 

Focusing leads to an average drift along the magnetic field lines, where the drift velocity is determined by the stable pitch-angle: $v_{\mathrm{drift}} = \mu_1 \mathrm{c}_0$. Spatial diffusion and pitch-angle diffusion without focusing have a mean drift velocity of zero at late times. The evolution of the mean drift term along the magnetic field line is shown in fig.~\ref{fig:meanpropagation}.

\section{Summary and Outlook}
\label{sec:summary}
This work introduces a more flexible version of and extension to the ensemble averaged description of cosmic ray transport with the open source propagation framework CRPropa. We explain how to derive the corresponding stochastic differential equations from a given (fractional) partial differential equation and show this examplarily for the focused transport equation (\ref{eq:focused_transport}) and different forms of the spatial transport equation (\ref{eq:transport}). We emphasized the difference between the derivation of the distribution function $f$ and the number density $n=fp^2$. 

The implementation is divided in three different modules (\texttt{DiffusionTensor}, \texttt{SDEParameter}, and \texttt{SDESolver}) to maximize flexibility and re-usability at the same time. In this way, we can, e.g., solve for the distribution function $f$ or number density $n$ by simply changing the \texttt{SDEParameter} module, corresponding to a single line in the python steering file. But also in the background most of the code is the same and only the method implementing the weighting term differs. This already allowed to easily extend the code to cover pitch angle diffusion and will be useful for future improvements (see below). 

Key parts of the code, namely the candidate splitting and the field line integration, have been validated again, ensuring that the results of the new implementation do not deviate from earlier less flexible versions of the software (see, e.g., \cite{Merten-etal-2017, Merten-etal-2018, Aerdker-etal-2024}). In addition to previous tests, we included a section discussing the advantages and disadvantages of using the \texttt{ObserverTimeEvolution}, starting individual simulation runs, and adding a continuous injection, which is another new feature (see sec.~\ref{ssec:cont_injection}). The quintessence of the analysis is: in using the time evolution observer approach earlier works might have underestimated the uncertainties, as multiple observations of the same candidate had been counted as statistically independent of each other. However, as can be seen in table \ref{tab:fit_parameters}, the best fit results do not differ significantly between the old and the updated more strict error estimation.

Furthermore, we extensively tested the newly available features of momentum and pitch angle diffusion. We could show that pitch angle diffusion is working as expected allowing to include this module in upcoming models of cosmic-ray re-acceleration, e.g., for low energetic Galactic cosmic rays. We verified the asymptotic behavior of isotropic pitch angle diffusion by comparing it to the analytic expectation and simulations of the corresponding spatial diffusion. In addition, we could show that including a constant focusing term $L=\mathrm{const.}$ leads to non trivial fixed points of the pitch angle's differential equation. When the stable fixed point falls into the domain $-1 < \mu < 1$, the particle ensemble will drift with an average velocity of $\langle v_\mathrm{drift}\rangle=v\mu_\mathrm{fixed}$ parallel to the magnetic field line direction. 

This work serves also as a reference with extended technical details for already published studies that used at least parts of the now rigorously explained software such as \cite{Aerdker-etal-2023, Aerdker-etal-2024, Effenberger-etal-2024} and will serve as the starting point for upcoming projects.

Parts of the described capabilities have been implemented in the publicly available version of CRPropa\footnote{See the, e.g., the GitHub page to get the latest public version: https://github.com/CRPropa/CRPropa3}. This includes a model for constant momentum diffusion \texttt{ConstantMomentumDiffusion} and the candidate splitting \texttt{CandidateSplitting}, as discussed in \cite{Aerdker-etal-2023}. Other parts of the code, such as time dependent advection fields will be included in the public version in the future.

\subsection{Outlook}
The flexibility of the developed stochastic differential equation solver will allow for many different extensions in the future, that will open up opportunities to model much more complex scenarios than those discussed here. This includes, e.g., drift terms coming from non-homogeneous magnetic field configurations, such as curvature or grad-B drifts. In combination with pitch angle diffusion parallel to the magnetic field line and a simple spatial perpendicular diffusion model it could be combined to a sophisticated model for transport of coronal mass ejections. 

Also an extension to model shock drift acceleration in addition and comparison to conventional Fermi first (DSA) and second (momentum diffusion) order acceleration is of interest. This would allow for more accurate descriptions of particle acceleration at oblique shocks as discussed in, e.g.~\cite{Isenberg-Jokipii-1979}. 

Spatially changing eigenvalues of the diffusion tensor lead to a drift term that is given by $\nabla\hat{\kappa}$ (see e.g., equ. \ref{eq:SDE_coeff_f-forward}). This will allow to model diffusion coefficients that change at the shock, as e.g.\ in \cite{Toptyghin-1980}.

Lastly, we want to emphasize, since the code is still based on CRPropa, that in principle all discussed transport aspects can be combined with all available interaction modules, e.g., describing synchrotron losses or interaction with ambient photon backgrounds.

\section*{Acknowledgments}

We acknowledge support from the Deutsche Forschungsgemeinschaft (DFG): this work was performed in the context of the DFG-funded Collaborative Research Center SFB1491 "Cosmic Interacting Matters - From Source to Signal" (grant no.~445990517).

\section*{Software}
The software used in this work will be made available on reasonable request to the corresponding author, if it has not yet been included in the public CRPropa repository.

The authors made use of the following software packages matplotlib \cite{Hunter-2007}, numpy \cite{Harris-etal-2020}, pandas \cite{McKinney-2010}, and jupyter \cite{Kluyver-etal-2016}.

\section*{CrediT Author Statement}
\textbf{Lukas Merten} Conceptualization, Methodology, Software, Validation, Formal Analysis, Writing - Original Draft, Writing - Review \& Editing, Visualization. \textbf{Sophie Aerdker} Conceptualization, Methodology, Software, Validation, Formal Analysis, Writing - Original Draft, Writing - Review \& Editing, Visualization.

\bibliography{references}
\bibliographystyle{elsarticle-num}

\appendix
\section{Field Line Integrator}
\label{apdx:fli}

\paragraph{Adaptive Step Refinement}
As explained above the field line integration uses a 4th order algorithm to calculate the position along the field line and in parallel a 5th order algorithm to calculate thr local truncation error
\begin{align}
    m = |\mathbf{r}_\mathrm{4th} - \mathbf{r}_\mathrm{5th}| \quad .
\end{align}
If the local truncation error is smaller than a user defined precision ($m< \xi\,\mathrm{kpc}$) the step is accepted. Otherwise the stochastic step $\sqrt{2\kappa_\parallel h}\eta_\parallel$ will be bisected until the above condition is met. Afterwards the field line integration will be repeated $2^{n}-1$ times, in case the step had to be bisected $n$ times.

The proposed next step is then
\begin{align}
    h_\mathrm{next} &= h^{-2n} \quad \textbf{if} n>1 \\
    h_\mathrm{next} &= 4h \quad \textbf{else} \quad .
\end{align}

See also \cite{Merten-etal-2017} for more details on this procedures. The validity is shown in the next paragraph.

\paragraph{Uncertainties}
To validate the field line integration algorithm the same tests as in \cite{Merten-etal-2017} have been performed. Using the path length along the field line, which is given by the sum of the stochastic step $L\mathrm{sim}=\sum_i \sqrt{2\kappa_\parallel h}\eta_\parallel$, and comparing it with the analytical solution
\begin{align}
    L = \int \left|\frac{\mathrm{d}\mathbf{r}_\mathrm{spiral}(z)}{\mathrm{d}z}\right|\,\mathrm{d}z
\end{align}
one can derive the analytically expected position on the field line $\mathbf{r}_\mathrm{ana}$. This can be used to calculate the first error $\Delta_1=|\mathbf{r}_\mathrm{ana} - \mathbf{r}_\mathrm{sim}|$, where $\mathbf{r}_\mathrm{sim}$ is the end position of the pseudo-particle. To judge how far the pseudo-particle deviated from it's original field line a second error $\Delta_2 = \min(\mathbf{r}_\mathrm{sim} - \mathbf{r}_\mathrm{spiral}$, where $\mathbf{r}_\mathrm{spiral}$ contains all points of the original field line. 

\begin{figure}
    \centering
    \includegraphics[trim={0, 0cm, 0, 0}, clip, width=0.8\linewidth]{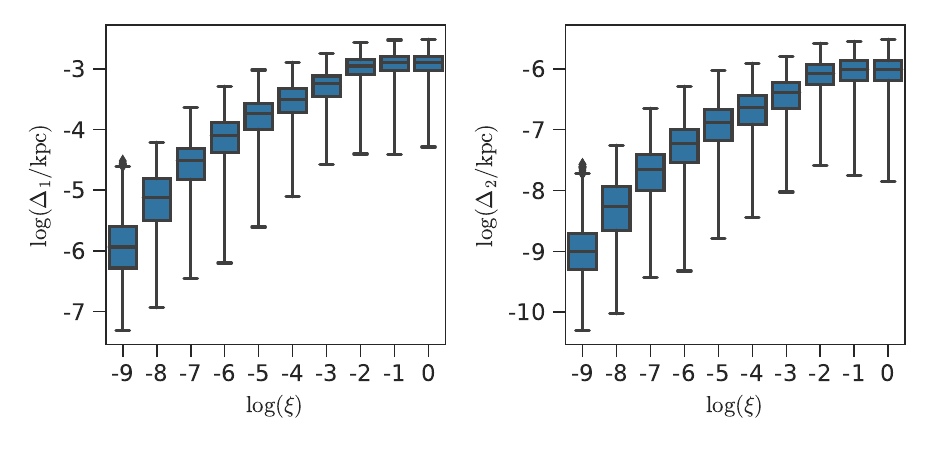}
    \caption{Errors of pure parallel diffusion along a spiral field line for different precisions $\xi$ of the field line integration algorithm. Left panel shows the deviation from the true position on the field line and the right panel shows the deviation from any point of the field line.}
    \label{fig:fli-error}
\end{figure}

The simulation parameters used for this test run are the same as in \cite{Merten-etal-2017}, $s=0.02\,\mathrm{kpc}$, $T_\mathrm{max}=100\,\mathrm{kpc}/c_0$, $h_\mathrm{min, max}=10^{-5}/1\,\mathrm{kpc}/c_0$. The tolerance or precision $\xi$ of the adaptive step refinement was varied as shown in fig.~\ref{fig:fli-error}.

\section{Transport Equation}
\label{asec:transportquation}
In comparison to the time forward FPE \ref{eq:FP-forward}, the time backward equation reads similarly:
\begin{align}
    \frac{\partial g(q_1, \dots, q_n, t)}{\partial t} = +  \sum_{i=1}^n A_i \frac{\partial g}{\partial x_i} + \frac{1}{2} \sum_{i, j} B_{ij} \frac{\partial^2 g}{\partial x_i \partial x_j} \quad. \label{eq:FP-backward}
\end{align}
Note, that for the time forward equation, the drift $\mathbf{A}$ and the diffusion term $\hat{B}$ are included in the derivatives and for the time backward case not. This will have an influence on the exact form of the SDE to be solved.

To derive the corresponding SDEs for a given transport equation, first is must be transformed into the proper Fokker Planck equations (FPEs) \ref{eq:FP-forward} or \ref{eq:FP-backward}. This is shown below for the distribution function $f$ as well as for the particle number density $n=fp^2$.

\subsection{Distribution function --- time backward}
\label{apdx:distribution-function}

Here, the FPE form of the transport equation for the time backward solution of the distribution function $f$ is derived. Starting from \ref{eq:transport}---omitting any source and loss term for now---, assuming again an isotropic particle distribution $f(\mathbf{r}, p)$. Furthermore, we use the Helmholtz decomposition of the velocity field $\mathbf{u}=\mathbf{v} + \mathbf{w}$, where $\mathbf{w} = -\nabla \phi$ and $\mathbf{v} = \nabla \times \mathbf{A}$.
\begin{align}
    \frac{\partial f}{\partial t} &= \nabla \cdot \left(\hat{\kappa} \nabla f - \mathbf{v}f \right) - \mathbf{w} \nabla f + \frac{p}{3}\nabla \cdot \mathbf{w} \frac{\partial f}{\partial p} + \frac{1}{p^2} \left( \frac{\partial}{\partial p} p^2 D \frac{\partial f}{\partial p} \right) \quad,  \label{eq:transportequation-f}
\end{align}
we start by transforming each summand into a form that is compatible with a FP transport equation.
\begin{align}
    \nabla\cdot(\hat{\kappa}\nabla f) &= (\nabla\hat{\kappa})\nabla f + \hat{\kappa}\nabla^2f \\
    \nabla\cdot(-\mathbf{v}f) &= - \mathbf{v}\nabla f \\
    \frac{1}{p^2}\left( \frac{\partial}{\partial p} D\frac{\partial f}{\partial p}\right) &= \left( \frac{2D}{p} + \frac{\partial D}{\partial p}\right) \frac{\partial f}{\partial p} + \frac{1}{2}(2D)\frac{\partial^2 f}{\partial p^2}, \quad , 
\end{align}
which gives the FP form as in:
\begin{align}
    \frac{\partial f}{\partial p} &= \frac{1}{2}(2\hat{\kappa})\nabla^2f +(\nabla\hat{\kappa} - \mathbf{u}) \nabla f \\
    &= \frac{1}{2}(2D)\frac{\partial^2 f}{\partial p^2} + \left(\frac{2D}{p} + \frac{\partial D}{\partial p} + \frac{p}{3}\nabla\cdot\mathbf{w} \right)\frac{\partial f}{\partial p} \quad.
\end{align}
The corresponding coefficients for the SDE read:
\begin{align}
    A_x = \nabla\hat{\kappa} + \mathbf{u} \quad &, \quad A_p = \left(\frac{\partial D}{\partial p} + \frac{2D}{p} + \frac{p}{3}\nabla \cdot \mathbf{w} \right) \quad ,\\
    B^2_x = 2\hat{\kappa} \quad &, \quad B_p^2 = 2D \quad .
\end{align}
The time-backward solution of the distribution function $f$ does not require any path weights. 

\subsection{Number Density --- time forward and backward}
\label{apdx:number-density}
Here, the FPE form of the transport equation for the particle number density $n=f p^2$ is derived. Starting from \ref{eq:transport}---omitting any source and loss term for now---, assuming again an isotropic particle distribution $f(\mathbf{r}, p)$. Furthermore, we use the Helmholtz decomposition of the velocity field $\mathbf{u}=\mathbf{v} + \mathbf{w}$, where $\mathbf{w} = -\nabla \phi$ and $\mathbf{v} = \nabla \times \mathbf{A}$.
\begin{align}
    \frac{\partial f}{\partial t} &= \nabla \cdot \left(\hat{\kappa} \nabla f - \mathbf{v}f \right) - \mathbf{w} \nabla f + \frac{p}{3}(\nabla \cdot \mathbf{w}) \frac{\partial f}{\partial p} + \frac{1}{p^2} \left( \frac{\partial}{\partial p} p^2 D \frac{\partial f}{\partial p} \right) \quad \left| \quad p^2\cdot \right . \\
    \frac{\partial n}{\partial t} &=  \nabla \cdot \left(\hat{\kappa} \nabla n - \mathbf{v}n \right) - \mathbf{w} \nabla n + \frac{p^3}{3}(\nabla \cdot \mathbf{w}) \frac{\partial f}{\partial p} + \left( \frac{\partial}{\partial p} p^2 D \frac{\partial f}{\partial p} \right) \quad , \label{eq:transport2}
\end{align}
where we still have mixed terms in \(f\) and $n$. Also this form cannot be easily transformed into stochastic differential equations. So now each summand is transformed in a form that is compatible with an FPE \ref{eq:FP-forward}
\begin{align}
    \nabla \cdot (\hat{\kappa} \nabla n -\mathbf{v} n) &= \nabla^2(\hat{\kappa} n) - \nabla \left[(\nabla \hat{\kappa})n + \mathbf{v}n \right] \quad , \label{eq:trafo_kappa}\\
    -\mathbf{w} \nabla n &= -\nabla \cdot (\mathbf{w} n) + (\nabla \cdot \mathbf{w}) n \quad , \label{eq:trafo_w}\\
    \frac{p^3}{3}(\nabla \cdot \mathbf{w}) \frac{\partial f}{\partial p} &= \frac{\partial}{\partial p} \left( \frac{p}{3} (\nabla \cdot \mathbf{w}) n \right) - (\nabla \cdot \mathbf{w}) n  \quad , \text{and} \label{eq:trafo_adiabatic}\\
    \frac{\partial}{\partial p} \left( D p^2 \frac{\partial f}{\partial p} \right) &= \frac{\partial^2}{\partial p^2}(Dn) - \frac{\partial}{\partial p} \left[\left(\frac{\partial D}{\partial p} + \frac{2D}{p}\right) n\right] \quad. \label{eq:trafo_D}
\end{align}
Inserting \ref{eq:trafo_kappa} to \ref{eq:trafo_D} into \ref{eq:transport2} and sorting the terms leads to:
\begin{align}
    \frac{\partial n}{\partial t} &= \frac{1}{2} \nabla^2 (2\hat{\kappa}n) - \nabla \left[\left(\nabla\hat{\kappa} + \mathbf{v} + \mathbf{w} \right) n \right]  \notag \\ 
    &+ \frac{1}{2} \frac{\partial^2}{\partial p^2} (2Dn) - \frac{\partial}{\partial p} \left[ \left(\frac{\partial D}{\partial p} + \frac{2D}{p} - \frac{p}{3}\nabla \cdot \mathbf{w} \right) n  \right] \notag \quad ,
\end{align}
which can easily be transformed into SDEs. The corresponding coefficients of the SDEs can be read of:
\begin{align}
    A_x = \left(\nabla\hat{\kappa} + \mathbf{v} + \mathbf{w} \right) \quad &, \quad A_p = \left(\frac{\partial D}{\partial p} + \frac{2D}{p} - \frac{p}{3}\nabla \cdot \mathbf{w} \right) \quad ,\\
    B^2_x = 2\hat{\kappa} \quad &, \quad B_p^2 = 2D \quad .
\end{align}

Source and loss terms as in the following equation:
\begin{align}
    \frac{\partial n}{\partial t} = - Ln + S \quad , \label{eq:source-sinks2}
\end{align}
cannot directly be included into the SDE but have to be treated by weighting the phase space elements or pseudo-particles in post-processing. Following the nomenclature of \textbf{Kopp} we call the factor introduced by the loss terms ($-Ln$) \emph{path weight} $\omega$ and the one coming from the sources or sinks ($S$) is called \emph{path amplitude} $w$. 
The path weight for an individual time step is given by $\omega_i = \exp(-L(\mathbf{r}_i, p_i, t_i)\Delta t_i)$. Since these weights are multiplicative this leads to:
\begin{align}
    \omega_j =  \exp \left(-\sum_{i=0}^j L(\mathbf{r}_i, p_i, t_i)\Delta t_i\right) \quad.
\end{align}
Analogously the path amplitude $w$ is increased (decreased) if a particle encounters a source (sink) region $w_i = w_{i-1} + S_i \omega_i \Delta t_i$, which can be written out as:
\begin{align}
    w_i = \sum_{j=0}^{i} S_j \Delta t_j  \exp \left(-\sum_{k=0}^j L(\mathbf{r}_k, p_k, t_k)\Delta t_k\right) \quad .
\end{align}

The derivation for the time backward equation is quite similar. Starting again from equation \ref{eq:transport2} we can derive the correct FP form by using:
\begin{align}
    \nabla \cdot (\hat{\kappa} \nabla n) &= (\nabla \hat{\kappa})\nabla n + \hat{\kappa} \nabla^2 n \quad, \\
    \nabla(-\mathbf{v}n) &= -\mathbf{v}\nabla n \quad , \\
    \frac{p^3}{3}\nabla\cdot \mathbf{w}\frac{\partial f}{\partial p} &= \frac{p^3}{3}\nabla\cdot \mathbf{w} \left( -\frac{2}{p^3}n +\frac{1}{p^2}\frac{\partial n}{\partial p}\right) \notag \\
    &= -\frac{2}{3} \ \nabla\cdot\mathbf{w} n + \frac{p}{3}\nabla\cdot\mathbf{w}\frac{\partial n}{\partial n} \quad , \\
    \frac{\partial}{\partial p}\left(p^2 D \frac{\partial f}{\partial f}\right) &= \left(2pD + p^2\frac{\partial D}{\partial p}\right) \cdot \left(-\frac{2}{p^3}n + \frac{1}{p^2}\frac{\partial n}{\partial p}\right) + p^2D \frac{\partial}{\partial p} \left( -\frac{2}{p^3}n = \frac{1}{p^2}\frac{\partial n}{\partial p}\right) \notag \\
    &= \frac{1}{2}(2D)\frac{\partial^2 n}{\partial p^2} - \left(\frac{2D}{p} - \frac{\partial D}{\partial p} \right) \frac{\partial n}{\partial p} - \left(-\frac{2D}{p^2} + \frac{2}{p}\frac{\partial D}{\partial p} \right) n \quad ,
\end{align}
where we used $\nabla\cdot\mathrm{v}=0$, $\partial_p f = -(2n)/p^3 + p^{-2}\partial_p n$. This leads to the FP form:
\begin{align}
    \frac{\partial n}{\partial t} &= \frac{1}{2} (2\hat{\kappa}) \nabla^2 + \left( \nabla \hat{\kappa} - \mathbf{u} \right) \nabla n \notag \\
    &+ \frac{1}{2}(2D) \frac{\partial^2 n}{\partial p^2} + \left(\frac{\partial D}{\partial p} - \frac{2D}{p} + \frac{p}{3} \nabla\cdot\mathbf{w}\right) \frac{\partial n}{\partial p} \notag \\
    &-\left(\frac{2}{3}\nabla\cdot\mathbf{w} + \frac{\partial}{\partial p}\frac{2D}{p} \right) n \quad , 
\end{align}
which leads to the following components of the SDE:
\begin{align}
    A_x = \nabla\hat{\kappa} - \mathbf{u} \quad &, \quad A_p = \frac{\partial D}{\partial p} - \frac{2D}{p} + \frac{p}{3} \nabla\cdot\mathbf{w} \quad ,\\
    B^2_x = 2\hat{\kappa} \quad &, \quad B_p^2 = 2D 
\end{align}
and an additional path weight which is an essential part to derive the correct distribution function
\begin{align}
    \omega_\mathrm{transp., i} = \exp\left[ \left(\frac{2}{3}\nabla\cdot\mathbf{w} - \frac{\partial}{\partial p}\frac{2D}{p} \right) \Delta t_i \right] \quad . \label{eq:transport_weight_n-bw}
\end{align}
Note, that the weights described in \ref{eq:transport_weight_n-bw} are relevant even when no losses from interactions are considered; making the solution of the forward and backward direction fundamentally different to solve.

Furthermore, it can be noted that the weights are symmetric in a sense, that they are required in the time-backward case for the number density $n$ and in the time-forward case of the distribution function.

\subsection{Pitch Angle Diffusion}
\label{apdx:pitchangle}

The focused transport equation as defined equ.\ \ref{eq:focused_transport} can be split into two equations for the pitch angle and the transport along the field line; $f(s, \mu, t)=f_1(\mu)f_2(s)$:

\begin{align}
    \frac{\partial f_2(s, t)}{\partial t} &= \frac{\partial}{\partial s}\left(\mu v f_2(s, t) \right) \\
    \frac{\partial f_2(\mu, t)}{\partial t} &= - \frac{\partial}{\partial \mu} \left[ \left( \frac{v}{2L}(1-\mu^2) + \frac{\partial D_{\mu\mu}}{\partial \mu} \right) f_1 \right] + \frac{1}{2} \frac{\partial^2}{\partial \mu^2} \left(2 D_{\mu\mu} f_1 \right) - \frac{\mu v}{L}f_1 \quad, 
\end{align}
where we used

\begin{align}
    \frac{\partial}{\partial \mu} \left(D_{\mu\mu} \frac{\partial f_1}{\partial \mu} \right) &= \frac{\partial^2}{\partial \mu^2} \left(D_{\mu\mu} f_1 \right) - \frac{\partial}{\partial \mu} \left(\frac{\partial D_{\mu\mu}}{\partial \mu} f_1 \right)  \quad \text{and}\\
    \frac{v}{2L}(1-\mu^2) \frac{\partial f_2}{\partial \mu} &= \frac{\partial}{\partial \mu} \left( \frac{v}{2L}(1-\mu^2)f_1\right) + \frac{v \mu}{L}f_1 \quad .
\end{align}
The corresponding coefficients for the SDE are:
\begin{align}
    A_s = \mu v \quad &, \quad B_s = 0 \\
    A_\mu = \frac{v}{2L}(1-\mu^2) + \frac{\partial D_{\mu\mu}}{\partial \mu} \quad &, \quad B_\mu^2 = 2D_{\mu\mu}
\end{align}
and the weighting term is given by:
\begin{align}
    w_{\mathrm{transp.}, i} = \exp\left( -\frac{\mu v}{L} \Delta t_i \right)
\end{align}

Momentum dependent terms can be added in the same way as shown for spatial diffusion in \ref{apdx:number-density} and \ref{apdx:distribution-function}.

\subsubsection{Boundary problem}
\label{sapdx:pitch_angle_boundary}

The stochastic differential equation for pitch angle diffusion is different compared to the one for spatial diffusion as the values of the pitch angle cosine $\mu$ are restricted to the open interval $\mu\in(-1, 1)$. As pitch angle scattering vanishes at the boundaries ($D_{\mu\mu}(\mu = \pm 1) = 0$) this is a numerical issue only. 

However, for any finite step size $h$, there is non vanishing possibility for a pseudo-particle's pitch angle to fall outside of the allowed range after a time integration step ($|\mu_{n+1}|>1$ for $|\mu_{n}|\approx 1$). Periodic boundary conditions could be implemented easily but this would lead to unphysical behaviour: A very small change in the pitch angle $\Delta \mu$ could lead to a flip of the pitch angle, e.g.\ from $\mu_n \approx 1$ to $\mu_{n+1}\approx -1$ and with that changing the propagation direction where only small angle scattering is expected. The alternative, that is also implemented here, is reflective boundary conditions for the pitch angle cosine domain. 

The numerical implementation is as follows:
\begin{align}
    m &= \mu_n + \Delta\mu\\
    \mu_{n+1} &= m \quad \text{for} \quad |m|\leq 1\\
    \mu_{n+1} &= \sign(m) - \sign(m)\cdot|m-\sign(m)| \quad ,
\end{align}
where $\sign(\pm|x|)=\pm 1$ is the signum function. For an illustration let's assume $\mu=0.9$ and $\Delta \mu = 0.15$, this will lead to $m=1.05$ and $\mu_{n+1}=1-1\cdot|1.05-1|=0.95$. Periodic boundary conditions instead would have lead to $\mu_{n+1, \mathrm{period.}}=-0.95$.

\end{document}